\documentclass[12pt,preprint]{emulateapj}
\usepackage[english]{babel}
\usepackage[T1]{fontenc}
\usepackage{amssymb}
\usepackage[]{graphicx}

\begin{document}

\title{Studying Frequency Relations of kHz QPOs for 4U 1636-53 and Sco X-1: Observations Confront Theories}

\author{Yong-Feng Lin\altaffilmark{1,2,3}, Martin Boutelier\altaffilmark{2,3}, Didier Barret\altaffilmark{2,3} and Shuang-Nan Zhang\altaffilmark{4,5}}
\altaffiltext{1}{Physics Department and Center for Astrophysics, Tsinghua University, Beijing 100084, China.}

\altaffiltext{2}{Universit{\'e} de Toulouse (UPS), 118 Route de Narbonne, 31062 Toulouse Cedex 9, France.}

\altaffiltext{3}{Centre National de la Recherche Scientifique, Centre d'Etude Spatiale des Rayonnements, UMR
5187, 9 av. du Colonel Roche, BP 44346, 31028 Toulouse Cedex 4, France.}

\altaffiltext{4}{Key Laboratory of Particle Astrophysics, Institute of High Energy Physics, Chinese Academy of
Sciences, P.O. Box 918-3, Beijing 100049, China.}

\altaffiltext{5}{Physics Department, University of Alabama in Huntsville, Huntsville, Alabama 35899, USA}

\email{zhangsn@ihep.ac.cn}

\begin{abstract}
By fitting the frequencies of simultaneous lower and upper kilohertz Quasi-Periodic Oscillations (kHz
QPOs) in two prototype neutron star QPO sources (4U~1636-536 and Sco X-1), we test the predictive power of all
currently proposed QPO models. Models predict either a linear, power-law or any other relationship between the two frequencies. We found that for plausible neutron star parameters (mass and angular
momentum), no model can satisfactorily reproduce the data, leading to very large chi-squared values in our
fittings. Both for 4U~1636-53 and Sco X-1, this is largely due to the fact that the data significantly
differ from a linear relationship. Some models perform relatively better but still have their own problems.
Such a detailed comparison of data with models shall enable to identify routes for improving those models
further.

\end{abstract}


\section{Introduction}

The launch of the X-ray timing satellite, \textsl{Rossi X-ray Timing Explorer} (RXTE), led to the discovery of
kilohertz Quasi Periodic Oscillations (kHz QPOs) of low mass X-ray binaries (LMXBs) in their X-ray
lightcurves. The frequencies of kHz QPOs range from a few hundreds to about one thousand Hz;
its time-scale corresponds to the dynamical time of the innermost regions of the accretion flow. Thus
such signals may carry crucial information about the central neutron star (NS), such as the mass, spin
frequency, angular momentum, radius, magnetic fields and so on. Usually the twin kHz QPOs appear simultaneously
and the lower and upper QPOs are almost directly proportional with each other \citep[see e.g.][]{van06}.

Various theoretical models have been proposed to account for the kHz QPO signals. Table \ref{tb:models} shows
all the present models we collect. Although each model achieves its success to a certain extent, the origin of
kHz QPOs is still highly debated. Furthermore many new models emerge in recent years, and systematic comparisons
among them have not been well studied.

\begin{table}[!hbp]\label{tb:models}
\begin{center}
\caption{Present models.}
\begin{tabular}{@{}cc@{}}
\hline
Models & Reference \\
\hline
Sonic-point and spin-resonance & [1, 2, 3] \\
Orbital resonance (3 models)& [4, 5] \\
Precession (3 models)       & [6, 7, 8] \\
Deformed-disk oscillation   & [9]   \\
`-1r, -2v' resonance        & [10, 11] \\
Higher-order nonlinearity   & [12]  \\
Tidal disruption            & [13]  \\
Rayleigh-Taylor gravity wave& [14, 15]  \\
MHD Alfv\'en wave oscillation & [16] \\
MHD                         & [17]  \\
Moving hot spots            & [18]  \\
\hline
\end{tabular}
\end{center}
[1] \citet{miller98}; [2] \citet{miller01}; [3] \citet{miller03}; [4] \citet{ab01}; [5] \citet{abet03}; [6]
\citet{stella99}; [7] \citet{bursa05}; [8] \citet{stuchlik07}; [9] \citet{kato01}; [10] \citet{torok07}; [11]
\citet{bakala08}; [12] \citet{muk09}; [13] \citet{cadez09}; [14] \citet{ot99}; [15] \citet{titar03}; [16]
\citet{zhang04}; [17] \citet{shi09}; [18] \citet{bahetti09}.
\end{table}

In view of this, we investigate systematically the predictive ability of the present kHz QPO models. We focus on
those that predict the frequency relation of twin kHz QPOs. The moving hot spots model is not included. This
model performs a 3D MHD simulations of the accretion around a NS. The simulations show that the moving hot spots
on the surface of a NS can develop oscillations in the lightcurves. However, this model does not provide any analytic relation between the twin QPOs \citep{bahetti09}.

In this work, we measure the frequency relations of the twin kHz QPOs for 4U 1636-53 and Sco X-1, then
fit the models with the measured results. We choose these two NS systems for several reasons. Firstly, both
of them have strong kHz QPOs over a wide frequency range; both of them have been observed more than 10 years
with RXTE. Thus the bias from the sample selection is minimized. Moreover, the different properties of
these two sources allow us to discuss the predictive ability of the models. They are typical Atoll and Z source,
respectively. The putative spin frequency for 4U 1636-53 is $581$ Hz \citep{strohmayer01, sm02},
whereas the spin frequency of Sco X-1 remains unknown.

In the following, we firstly describe the data reduction procedure. Then we fit the frequency relations with all
the available models. The predictions of NS properties in each model will be presented. Finally we discuss and
conclude our investigation results.

\section{Data Analysis}

We have retrieved all the public archival data of the two sources with the Proportional Counter Array (PCA) on
board \emph{RXTE}. The observation time is from Feb. 28th, 1996 to Sep. 25th, 2007 for 4U 1636-53, and from May
5th, 1996 to Feb. 4th, 2006 for Sco X-1.

For 4U 1636-53, we use the event mode data for 1156 ObsIDs, with time resolution better than 256 $\mu$s and an
energy band of 2-40 keV. With the similar analysis procedure in \citet{barret06} and \citet{bou09m}, the PDS as
well as the QPO parameters (peak frequency, width and amplitude) in each ObsID are obtained. For the ObsIDs with
PDS containing the lower QPO, we track its time evolution in every 128 s. Following that in \citet{barret05},
all the 128 s PDS are aligned in every 30 Hz interval of the lower QPO with the shift-and-add technique
\citep{mendez98}. Then we search for twin QPOs in each interval. For the ObsID with PDS containing only the
upper QPOs, we directly align the PDS of each ObsID in every 30 Hz interval of upper QPO. Again the twin QPOs in
each interval are searched. Finally, the two parts of results are combined and we obtain the frequency relation.

For Sco X-1, we analyze the Generic Binned mode data in 187 ObsIDs with time resolution better than 256 $\mu$s
and an energy band of 2-40 keV. Considering the effects of the deadtime, we use a model of two
Lorentzians plus a powerlaw to fit each PDS. The powerlaw component denotes the deadtime-modified Poisson noise;
the Lorentzians account for the contribution of the twin kHz QPOs. We then apply the shift-and-add technique to
the ObsID averaged PDS as described above on the upper QPOs' frequency, because the span and significance of the
upper QPOs are larger than that of the lower QPOs. The interval of shift-and-add is 50 Hz. Similar to the result
in \citet{van00}, our frequency relation shows some subtle structure when the lower QPO is around 800 Hz.

\section{Comparisons between Models and Data}

In the following, we restrict the NS parameters $M\in[1.4, 2.4]$ $M_\odot$ and $j\in[0, 0.3] $
($j\equiv{Jc/GM^2}$) in our fittings, corresponding to reasonable equations of state (EOS) of NSs (Lattimer \&
Prakash 2007), where $M$ and $j$ are the mass and dimensionless angular momentum parameter of a NS,
respectively. The fitting results are summarized in Table 2. For some models, the best fitting
values of $M$ and $j$ approach the upper or lower limits. In each of these cases, we made an extended fitting to
relax the limits to $M\in[1.0, 4.0]$ $M_\odot$ and $j\in[0, 0.5] $; these results are presented in Table 3 and
discussed in Section \ref{se:con}.

\subsection{Comparison with the sonic-point and spin-resonance model}

The sonic-point and spin-resonance model \citep{miller98, miller01, miller03} attributes the formation of the
twin kHz QPOs to the interaction between the orbital motion of the flow and the NS rotation. The interaction
happens at the `sonic point' where the radial inflow becomes supersonic. In the frame of the model, the X-ray
source is a NS with a surface magnetic field about ${10^7\sim10^{10}}$ G and a spin of a few hundreds Hz, which
accretes gas via a Keplerian disk. At the sonic point $r_{\rm sp}$, some of the accreting gas is channeled by
the magnetic field and then impacts the NS surface to produce the lower QPO. Some remains in clumps with the
Keplerian disk flow, producing the upper QPO. Therefore the upper frequency $\nu_2$ is the Keplerian frequency
$\nu_{\rm K}$ at $r_{\rm sp}$; the lower one $\nu_1$ is the beat frequency between $\nu_{\rm K}$ and the NS spin
$\nu_{\rm s}$, i.e. $\nu_1\approx\nu_{\rm B}=\nu_{\rm K}-\nu_{\rm s}$. The first version of the model
\citep{miller98} leads to a constant peak separation $\Delta\nu$, close to $\nu_{\rm s}$. The second version
\citep{miller01} introduced inward drifts of gas to make $\Delta\nu$ dependent on $\nu_1$ (or $\nu_2$). The
inward drifts make $\nu_1$ greater than $\nu_{\rm B}$ and $\nu_2$ less than $\nu_{\rm K}$ for a prograde gas
flow,
\begin{equation}\label{SPBM1}
\nu_1\approx{\nu_{\rm B}}/(1-\upsilon_{\rm cl}/\upsilon_{\rm g}),
\end{equation}
\begin{equation}
\nu_2\approx{\nu_{\rm K}}(1-\frac{1}{2}\upsilon_{\rm cl}/\upsilon_{\rm g}), \label{SPBM2}
\end{equation}
where $\upsilon_{\rm cl}$ is the inward radial velocity of clumps near $r_{\rm sp}$, $\upsilon_{\rm g}$ the
characteristic inward radial velocity of gas. $\upsilon_{\rm cl}$ and $\upsilon_{\rm g}$ are supposed to be
approximately constant during the lifetime of a clump, and $\upsilon_{\rm cl}\ll{\upsilon_{\rm g}}$.

\citet{miller03} proposed the third version to explain that the frequency separation is close to
$\nu_{\rm spin}$ in some stars but close to $\nu_{\rm spin}/2$ in others. The upper QPO is likewise close to the
Keplerian frequency $\nu_{\rm K}$ at $r_{\rm sp}$. They showed that magnetic and radiation fields rotating with
the star will preferentially excite vertical motions in the disk at the `spin-resonance' radius $r_{\rm sr}$
where $\nu_{\rm K}-\nu_{\rm s}$ is equal to the vertical epicyclic frequency. There are two cases in this model.
Case 1 supposes that the flow at $r_{\rm sr}$ is relatively smooth, then the vertical motions excited at $r_{\rm
sr}$ modulate the X-ray flux at $\nu_1\approx \nu_2- \nu_{\rm s}$. Case 1 is fully compatible with the second
version \citep{miller01}. Case 2 assumes that the flow at $r_{\rm sr}$ is highly clumped. In this case, the
vertical motions excited at $r_{\rm sr}$ modulate the X-ray flux at $\nu_1\approx \nu_2-\nu_{\rm s}/2$.

\begin{figure}[!htp]
\includegraphics[width=80mm]{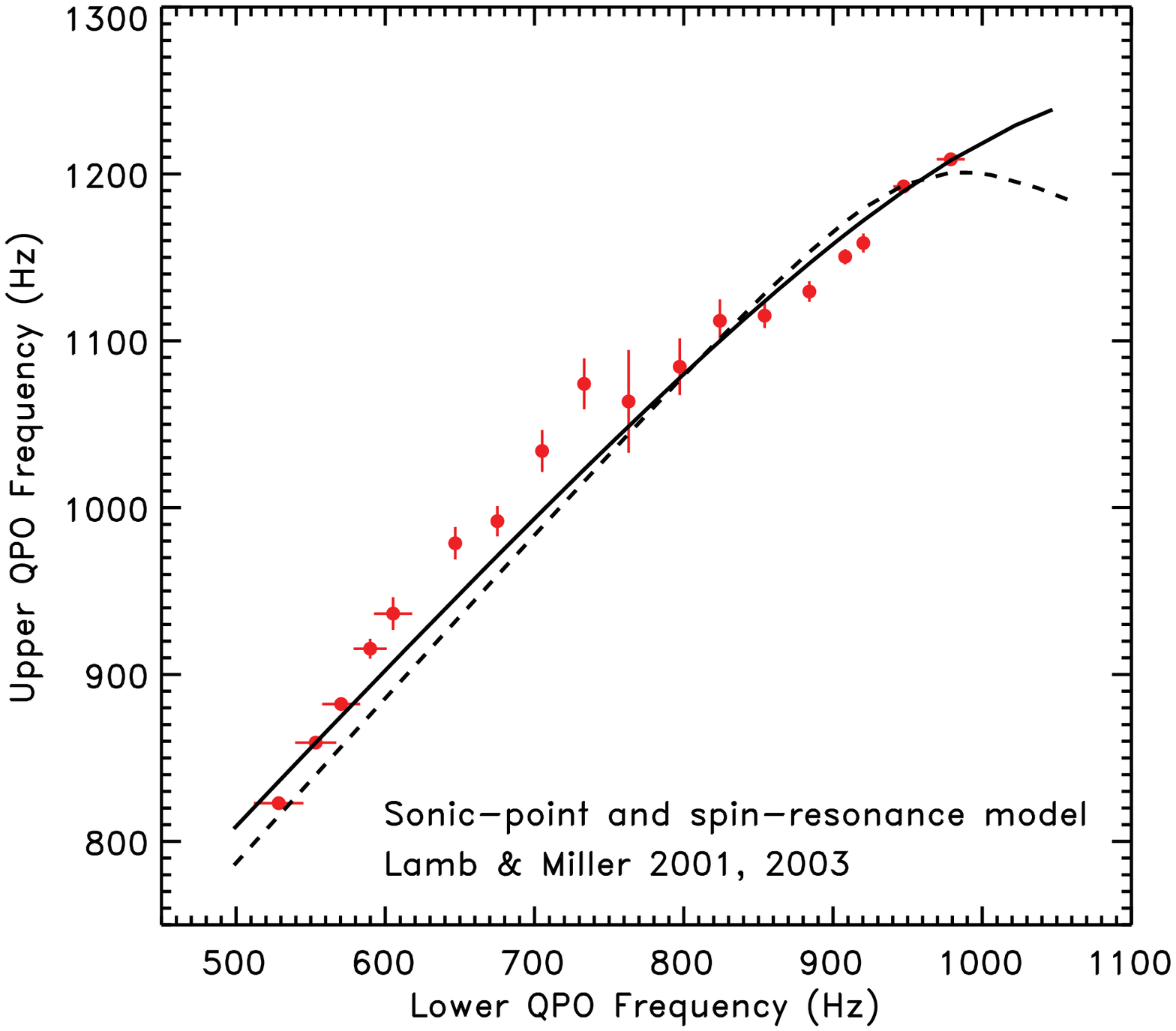}
\includegraphics[width=80mm]{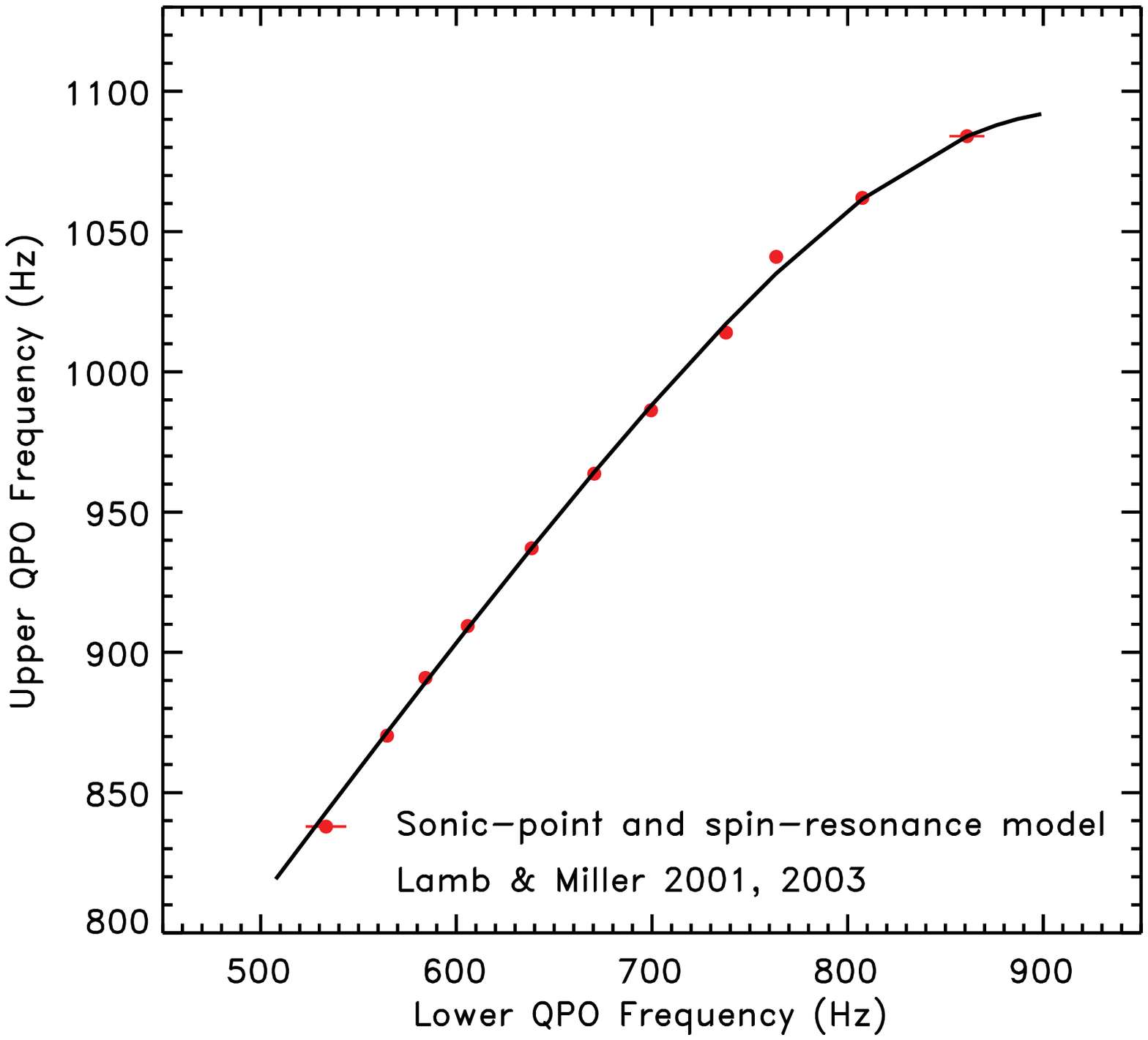}
\caption{Fitting results to the sonic-point and spin-resonance model for 4U 1636-53 (top) and Sco X-1 (bottom).
The data points with error bars are the observed frequency relations and the curves represent the model
predictions. In the top panel, the solid curve represents the best fitting result, the dashed one shows the
fitting result by fixing $\nu_{\rm s}=581$ Hz.}\label{fig:SPBM}
\end{figure}

Figure \ref{fig:SPBM} (top panel) displays the fitting result for 4U 1636-53. We set $\nu_1\approx
\nu_2-\nu_{\rm s}/2$ as 4U 1636-53 belongs to the second case in \citet{miller03}. The ratio $\upsilon_{\rm
cl}/\upsilon_{\rm g}$ is represented by a free parameter, the torque coefficient $c_N$ \citep[see][for
detail]{miller01}. Our fitting result is $M=1.545$ $M_\odot$, $\nu_{\rm s}=652\pm 5$ Hz, $c_N=0.00274$. The
fitting does not give a spin frequency close to $581$ Hz. Moreover, our measured $\Delta\nu$ ranges from $220$
to $340$ Hz, which could be smaller or larger than $\nu_{\rm s}/2$ ($290.5$ Hz). In fact, such behavior of
$\Delta\nu$ is already shown in \citet{jonker02}. Though successful in explaining a changing $\Delta\nu$ close
to half of the spin frequency, the latest version predicts that $\Delta\nu$ is always smaller (larger) than
$\nu_{\rm s}/2$ for a prograde (retrograde) flow. When $\nu_1$ is below about $800$ Hz,
$\Delta\nu$ is larger than $290.5$ Hz; otherwise, it is smaller than $290.5$ Hz.  Finally one can notice that
the fitting result by fixing $\nu_{\rm s}=581$ Hz shows larger deviations from the data points.

The fitting result for Sco X-1 is shown in Figure \ref{fig:SPBM} (bottom panel), giving $M=1.80$
$M_\odot$, $\nu_{\rm s}=329.5$ Hz (or $\nu_{\rm s}=659$ Hz) and $c_N=0.00216$, depending on which case we
choose. Comparing with the fitting result in \citet{miller01} and \citet{miller01}, we get a larger $M$ and a
slightly smaller $\nu_{\rm s}$. The reduced $\chi^2$ is slightly larger, partly due to our more precise results
with smaller error bars.

\subsection{Comparisons with orbital resonance models}\label{se:orb_res}

\citet{ab01}, \citet{abet03} introduced several kHz QPO models based on the idea of the resonances between the
radial and vertical frequencies in orbital motion.

\subsubsection{The $2:3$ parametric resonance model}

A parametric resonance instability occurs near $\omega_r= 2\omega_\theta/n$ for $n = 1, 2, 3, \cdots$ in an
oscillator that obeys a Mathieu-type equation of motion,
\begin{equation}
\delta\ddot{\theta}+\omega^2_\theta[1+h\cos(\omega_rt)]\delta\theta=0\ ,
\end{equation}
where $\delta\theta$ is the small deviation of elevation $\theta$, the dot denotes the time derivative, $h$ is a
known constant. $\omega_r=2\pi\nu_r$ and $\omega_\theta=2\pi\nu_\theta$, where $\nu_r$ and $\nu_\theta$ are the
radial and vertical epicyclic frequency, respectively \citep{abet03}.

The model predicts $\nu_r:\nu_\theta=2:n$. When $n$ has the smallest possible value, the strongest resonance is
excited. Since $\nu_r<\nu_\theta$, the smallest possible value for resonance is $n=3$ , meaning that
$\nu_r:\nu_\theta=2:3$. Simply supposing $\nu_1= \nu_r$ and $\nu_2= \nu_\theta$ \citep{ab02}, one can infer the
$2:3$ ratio of the twin kHz QPO peak frequencies. The excitation of the resonance has been studied with
numerical simulations \citep{abet03} and an analytic method \citep{rebu04}.

\begin{figure}[!htp]
\includegraphics[width=80mm]{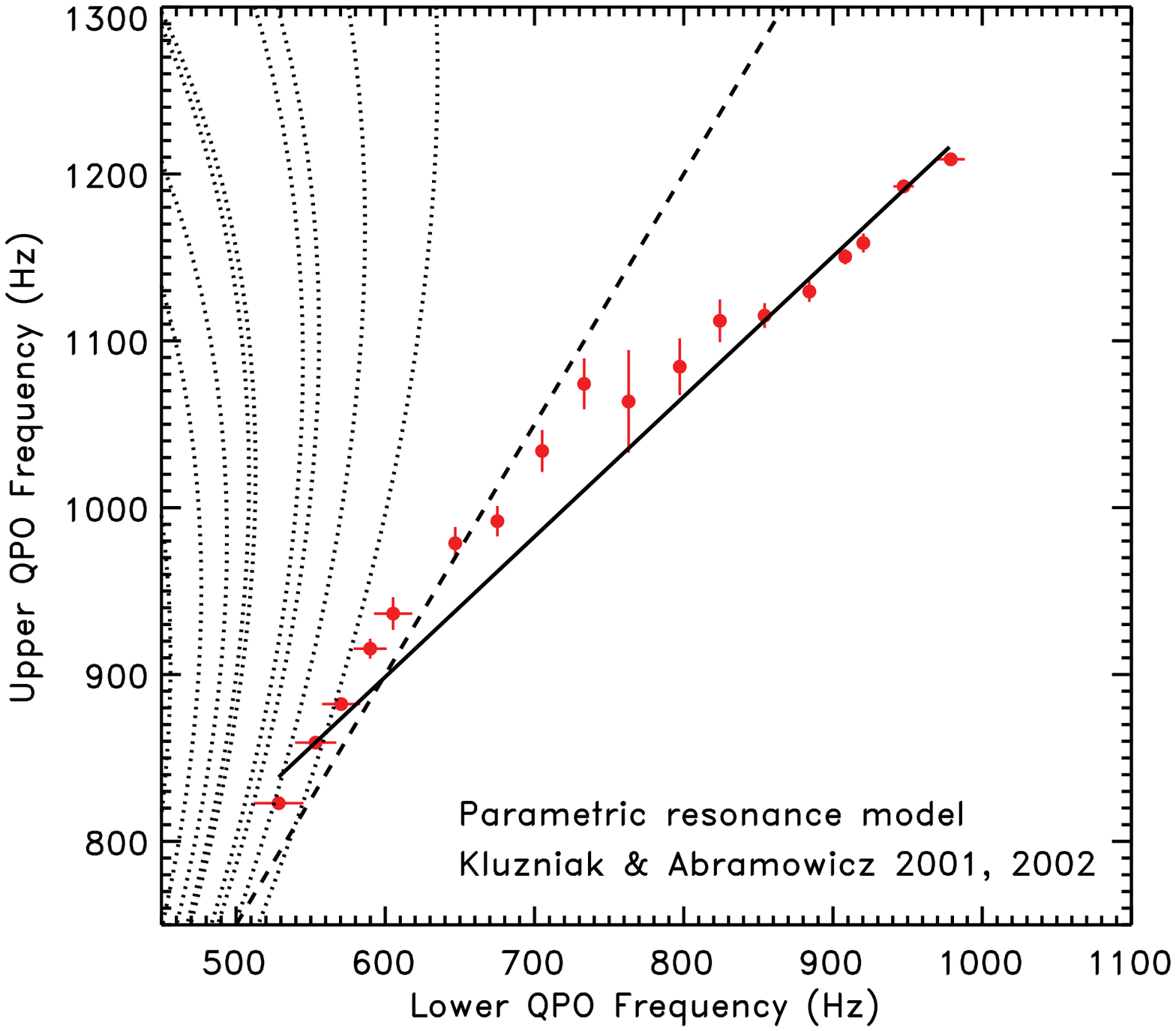}
\includegraphics[width=80mm]{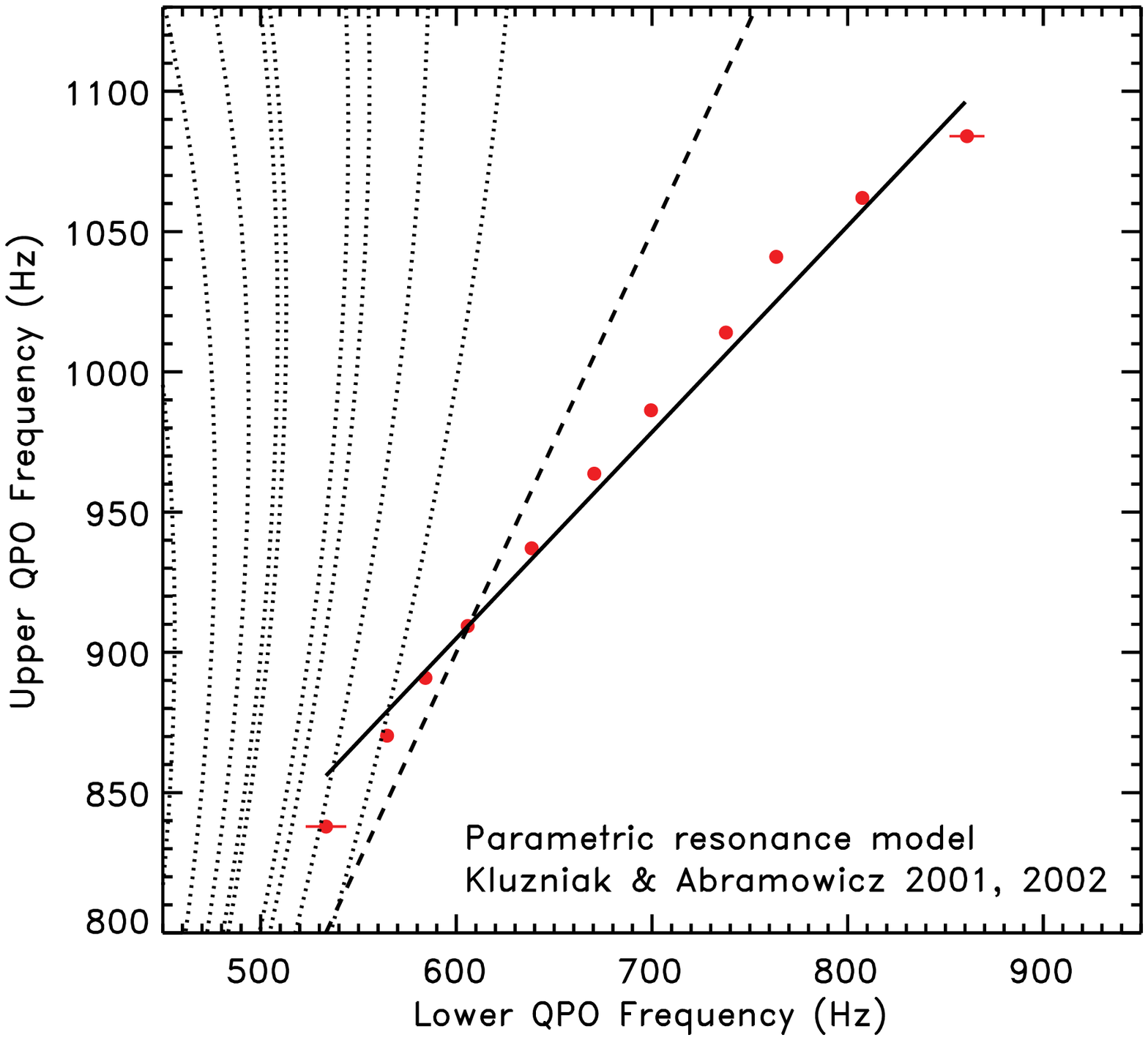}
\caption{Fitting results to the parametric resonance model for 4U 1636-53 (top) and Sco X-1(bottom). The dashed
curve represents the model with $\nu_1:\nu_2=2:3$. The dotted curves denote the simple assumption ($\nu_2=
\nu_\theta$, $\nu_1= \nu_r$) with different $M$ and $j$. The solid curve shows a linear frequency relation
according to \citet{abet05}. }\label{fig:ab_res}
\end{figure}

The behavior of the frequency relation in the parametric resonance model is shown in Figure \ref{fig:ab_res}, in
comparison to our measured data. At first, one can notice that the linear $2:3$ frequency relation (dashed) is
not in agreement with the observations. Then we also notice that the model with $\nu_2= \nu_\theta$ and $\nu_1=
\nu_r$ deviates from data much more (dotted); it is even unable to follow the basic downward-bending track of
the data points. Actually the assumption $\nu_1= \nu_r$ is inappropriate for a NS. Under the condition of
$M\in[1.4, 2.4]$ $M_\odot$ and $j\in[0, 0.3]$, theoretical $\nu_r$ has a maximum value about $635$ Hz when
$M=1.4$ $M_\odot$, $j=0.3$ \citep[see e.g.][the equations of orbital frequencies]{stella99}. However, the
measured $\nu_1$ reaches as large as $800\sim1000$ Hz. As shown in the figure, the rightmost dotted curve
represents the predicted frequency relation with $M=1.4$ $M_\odot$, $j=0.3$ and the leftmost one with $M=2.4$
$M_\odot$, $j=0$. The predicted curves with other values of $M$ and $j$ lie between them. Finally a linear
frequency relation, i.e. $\nu_2=\nu_\theta$ and $\nu_2=k\nu_1+b$, is proposed in \citet{abet05}. We use it to
fit the data and get $k=0.840$, $b=395$ Hz for 4U 1636-53, and $k=0.805$, $b=422$ Hz for Sco X-1, respectively.
Then we get the ratio of $\nu_1$ to $\nu_2$ for these two sources to be $0.725$ and $0.683$, respectively, close
to but higher than $2:3$. Adopting a function as $\nu_2=k*\nu_1+b$, \citet{belloni05} also concluded the
similar ratios. We will discuss the ratios in Section \ref{se:con}. Here we just show that the linear fit is
naturally disfavored by the data points with apparent non-linearity.

\subsubsection{The forced $1:2$ and $1:3$ resonance model}\label{se:forced}

In the numerical simulations of oscillations of a perfect fluid torus \citep{abet03}, there is an evident
resonant forcing of vertical oscillations. The forcing is caused by the radial oscillations through a pressure
coupling. This result supports another possible resonance model \citep{abk01, ab04}. In the model, the
resonances occur in a forced non-linear oscillator,
\begin{equation}
\delta\ddot{\theta}+\omega^2_\theta\delta\theta+[{\rm non\ linear\ terms\ in} \
\delta\theta]=h(r)\cos(\omega_rt), \omega_\theta=n\omega_r
\end{equation}
here again $\omega_r=2\pi\nu_r$ and $\omega_\theta=2\pi\nu_\theta$.

This model predicts that one of the combination frequencies, i.e. $\nu_-=\nu_\theta-\nu_r$ and
$\nu_+=\nu_\theta+\nu_r$, has a $2:3$ ratio to the vertical frequency. For $n=2$, the forced epicyclic resonance
$\nu_r:\nu_\theta=1:2$,
\begin{equation}
\nu_1=\nu_\theta,   \qquad   \nu_2=\nu_+,
\end{equation}
and for $n=3$, the forced epicyclic resonance $\nu_r:\nu_\theta=1:3$,
\begin{equation}
\nu_1=\nu_-,   \qquad   \nu_2=\nu_\theta.
\end{equation}

\begin{figure}[!htp]
\includegraphics[width=80mm]{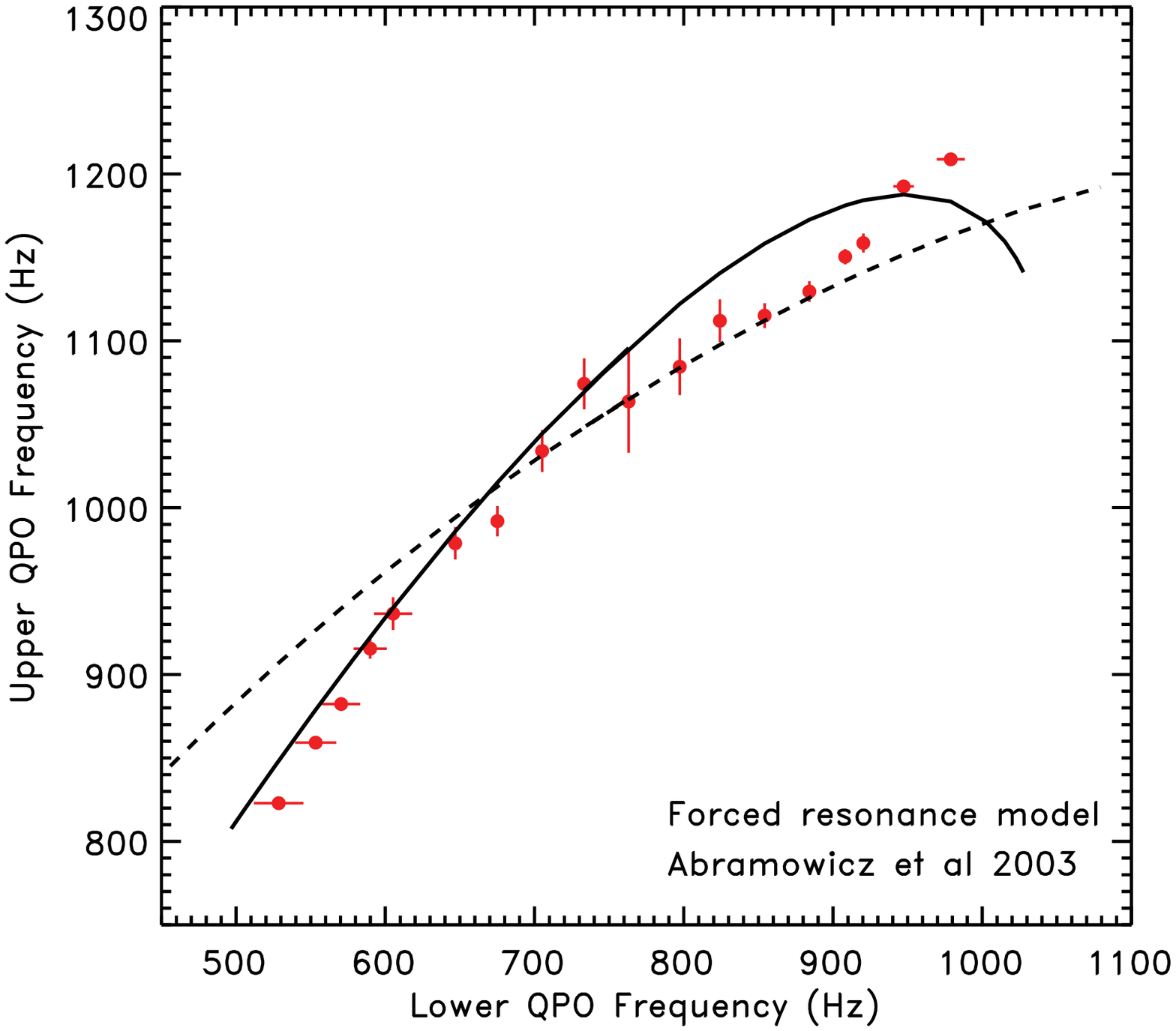}
\includegraphics[width=80mm]{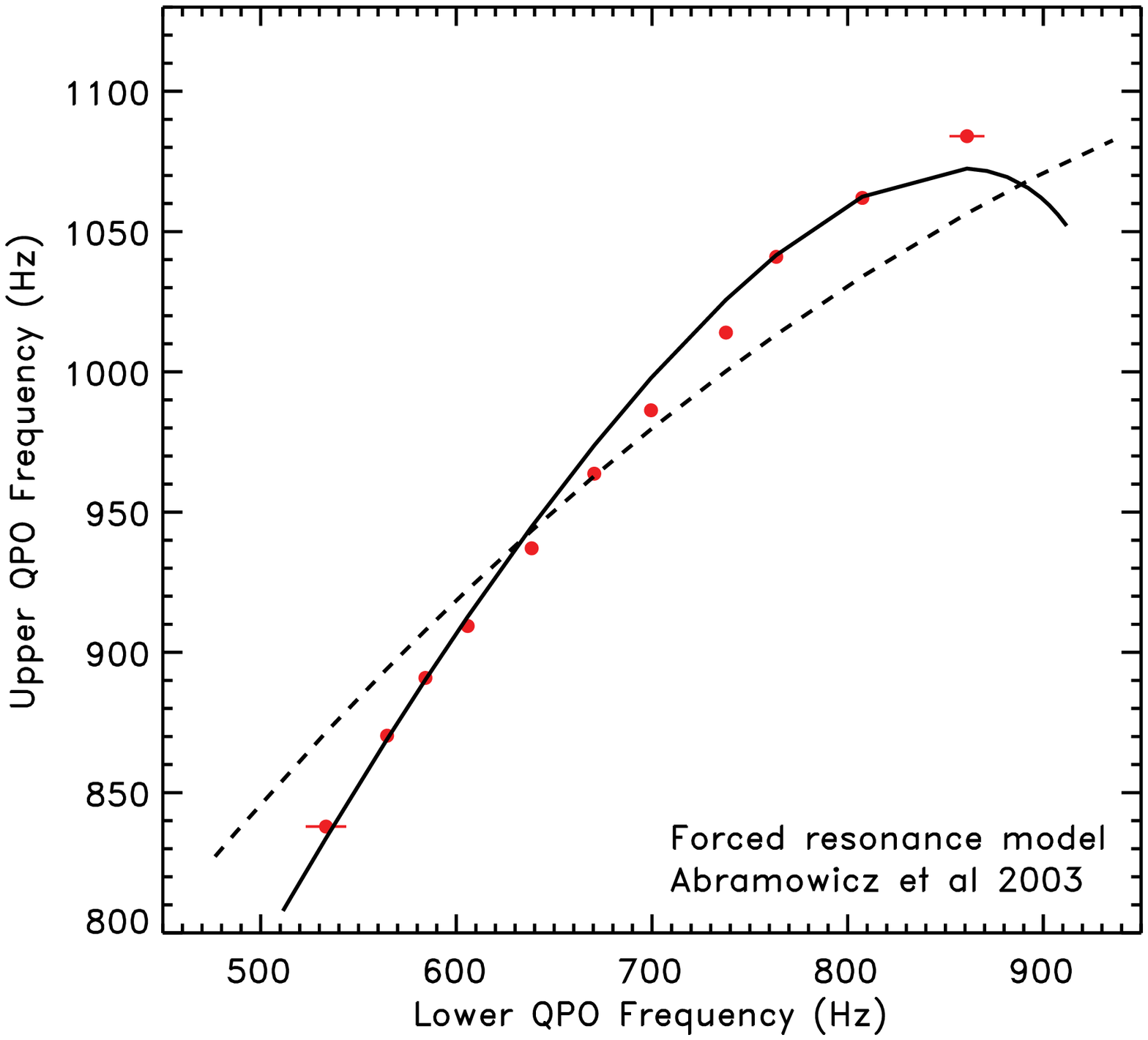}
\caption{The fitting results to the forced resonance models for 4U 1636-53 (top) and Sco X-1(bottom). The solid
curve and the dashed one show the forced $1:2$ and $1:3$ resonance model, respectively. }\label{fig:forced}
\end{figure}

We fit the observed QPO frequency relations with the forced resonance models in Figure \ref{fig:forced}. The
forced $1:2$ model predicts that $\nu_2$ climbs up to a maximum value at $\nu_1\approx950$ Hz, then decreases
rapidly. However, our analysis results for the two LMXBs do not show the trend that $\nu_2$ should decrease as
$\nu_1$ increases. Regarding the forced $1:3$ model, it cannot adequately describe the data points, especially
at the high and low frequencies. The $\nu_2$ predicted is higher than that observed at low frequencies, but
lower than that observed at high frequencies. It means that the observed $\Delta\nu$ does not decrease so
sharply as the model predicts. In addition, these two forced models give very large reduced $\chi^2$.

\subsection{Comparisons with precession models}\label{se:precession}

This section investigates the predictive ability of three precession models, namely, relativistic precession
model, vertical precession model and total precession model. In these precession models, QPOs can be excited by
various resonances with the precession frequencies and orbital frequencies under certain conditions, such as inhomogeneities orbiting the inner disk boundary \citep{stella01}.

For the well-known relativistic precession model \citep{stella99}, the upper QPO $\nu_2$ is assumed to be the
azimuthal frequency $\nu_\phi$, the lower QPO $\nu_1$ is expressed as the relativistic periastron precession
frequency,
\begin{equation}\label{eq:peri}
\nu_1=\nu_\phi-\nu_r,
\end{equation}
\begin{equation}
\nu_2=\nu_\phi.
\end{equation}

In the vertical precession model \citep{bursa05}, $\nu_1$ is same as that in Eq.~(\ref{eq:peri}); $\nu_2$ is
hypothesized as $\nu_\theta$.

In the total precession model \citep{stuchlik07}, $\nu_1$ is the total precession frequency, and $\nu_2$ is
ascribed to the Keplerian frequency $\nu_{\rm K}$ (or the vertical frequency $\nu_\theta$),
\begin{equation}
\nu_1=\nu_\theta-\nu_r
\end{equation}
\begin{equation}
\nu_2=\nu_{\rm K}   \qquad or \qquad \nu_2=\nu_\theta.
\end{equation}

\begin{figure}[!htp]
\includegraphics[width=80mm]{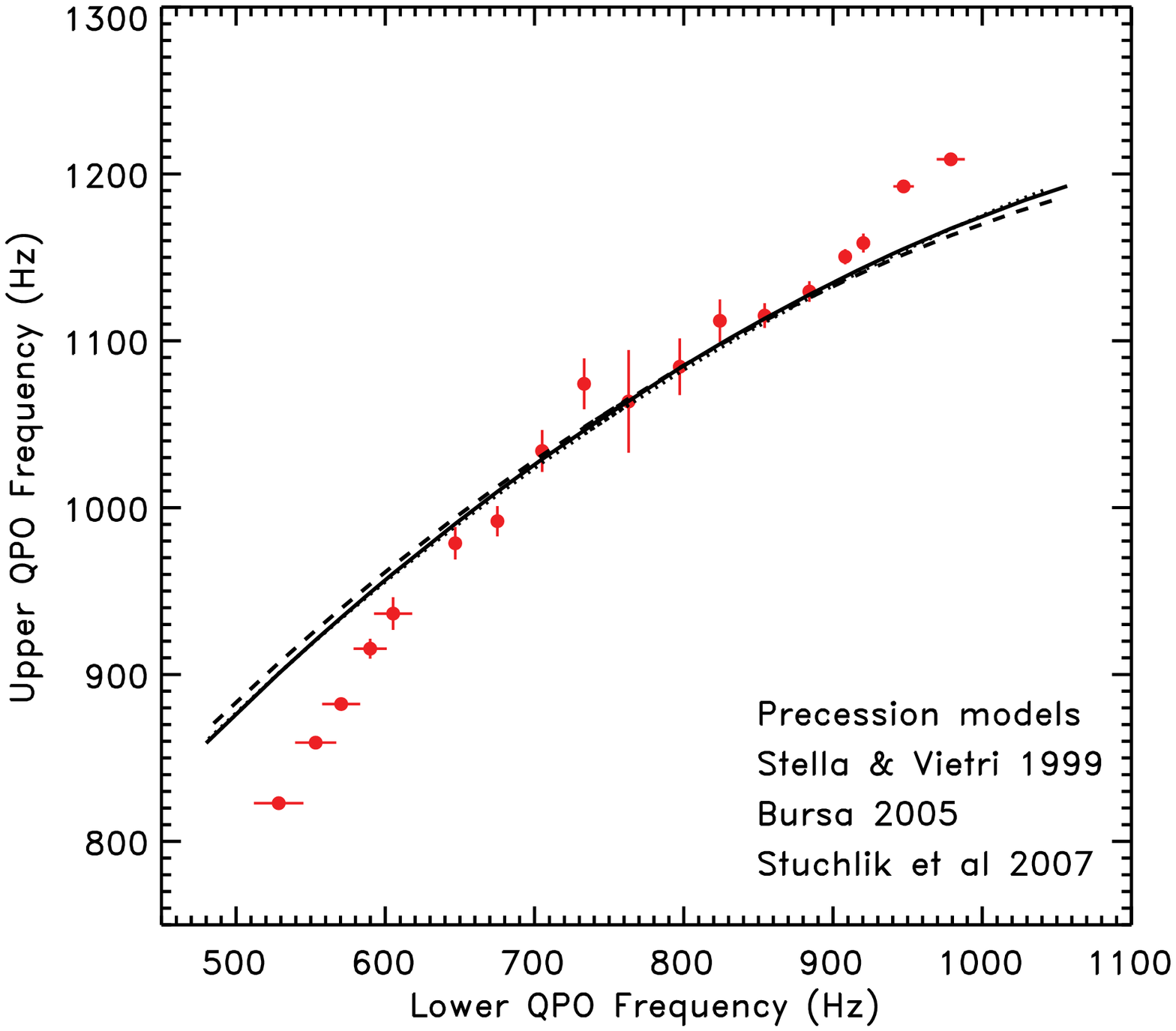}
\includegraphics[width=80mm]{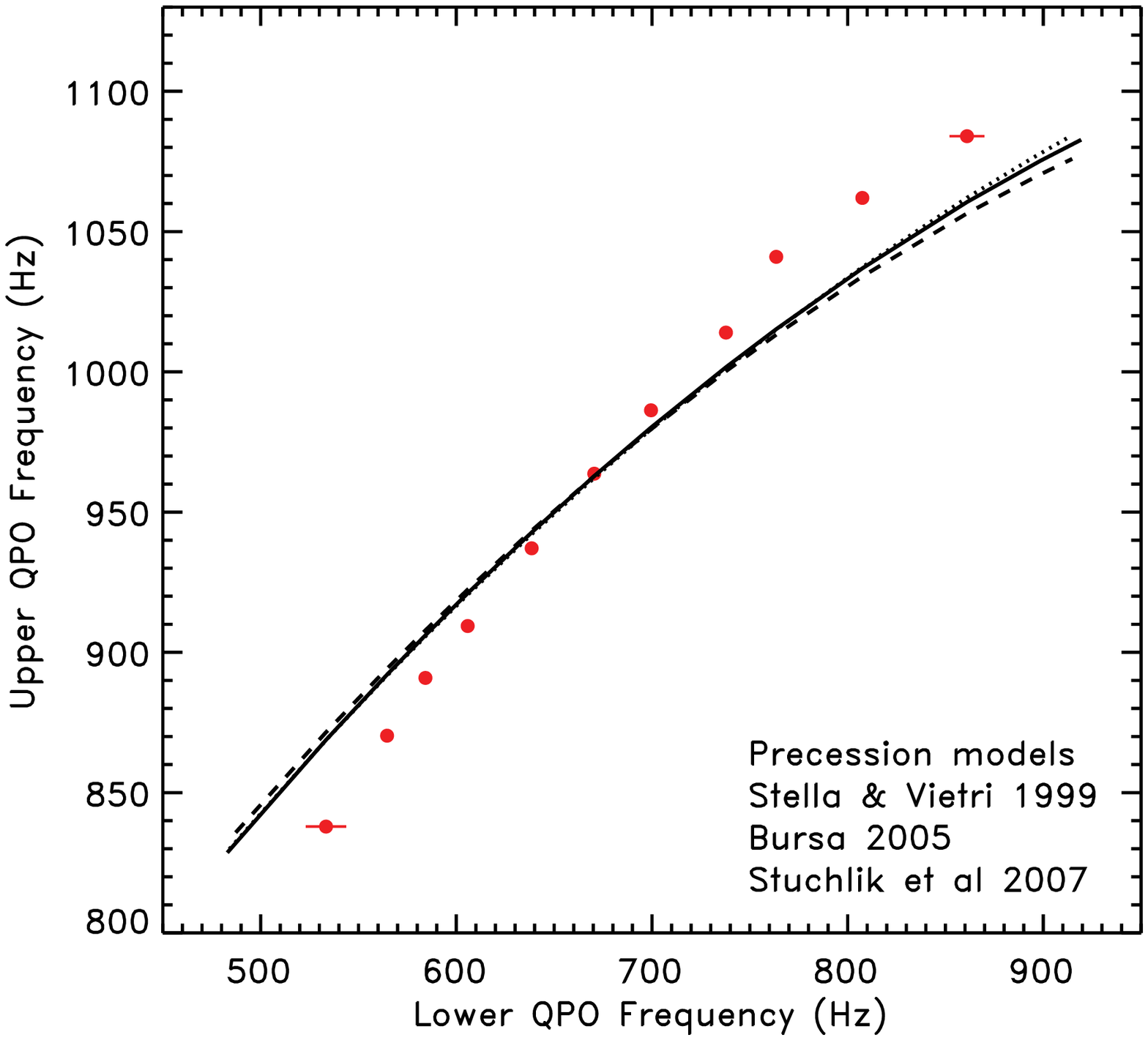}
\caption{The fitting results to the precession models for 4U 1636-53 (top) and Sco X-1 (bottom). The solid,
dotted and dashed curves represent the predictions of the relativistic, vertical and total precession models,
respectively. The three curves almost overlap, especially the relativistic and vertical models.
}\label{fig:precession}
\end{figure}

Figure \ref{fig:precession} illustrates the comparison of the precession models with the observed data.  Notice
that the models almost overlap and have the same deviation as the forced $1:3$ resonance model: $\nu_2$ is
predicted too low at high frequencies and too high at low frequencies. Hence the deviation from the observations
increases significantly at high and low frequencies. In addition, the relativistic and vertical precession
models give large $M$ and $j$. Though the total precession model give smaller $M$ and $j$, the fit has largest
$\chi^2$ among the precession models.

\subsection{Comparisons with disk oscillation models}

The resonances between specific modes in an accretion disk are also studied for exciting the observed kHz QPOs.
In this section, we investigate two models of this kind.

\subsubsection{The deformed-disk oscillation model}

\citet{kato01} brought forward the deformed-disk resonance model. kHz QPOs are excited by a horizontal resonance
in a deformed (warped or eccentric) disk under inviscid and adiabatic perturbations. The perturbations vary as
$\exp[i(\omega t-m\phi)]$, where $\omega$ is the frequency of the perturbations and $m$ ($= 0, 1, 2, \cdots$)
denotes the number of arms in the azimuthal direction. Various modes of perturbations are considered in a series
of their works \citep{kato03, kato05, kato09}. In the model, QPOs are inertial-acoustic oscillations (p-mode)
and the gravity oscillations (g-mode), or their combination. The twin kHz QPOs are,
\begin{equation}
\nu_1 = 2(\nu_{\rm K} - \nu_r)
\end{equation}
\begin{equation}
\nu_2= 2\nu_{\rm K} -\nu_r.
\end{equation}

\begin{figure}[!htp]
\includegraphics[width=80mm]{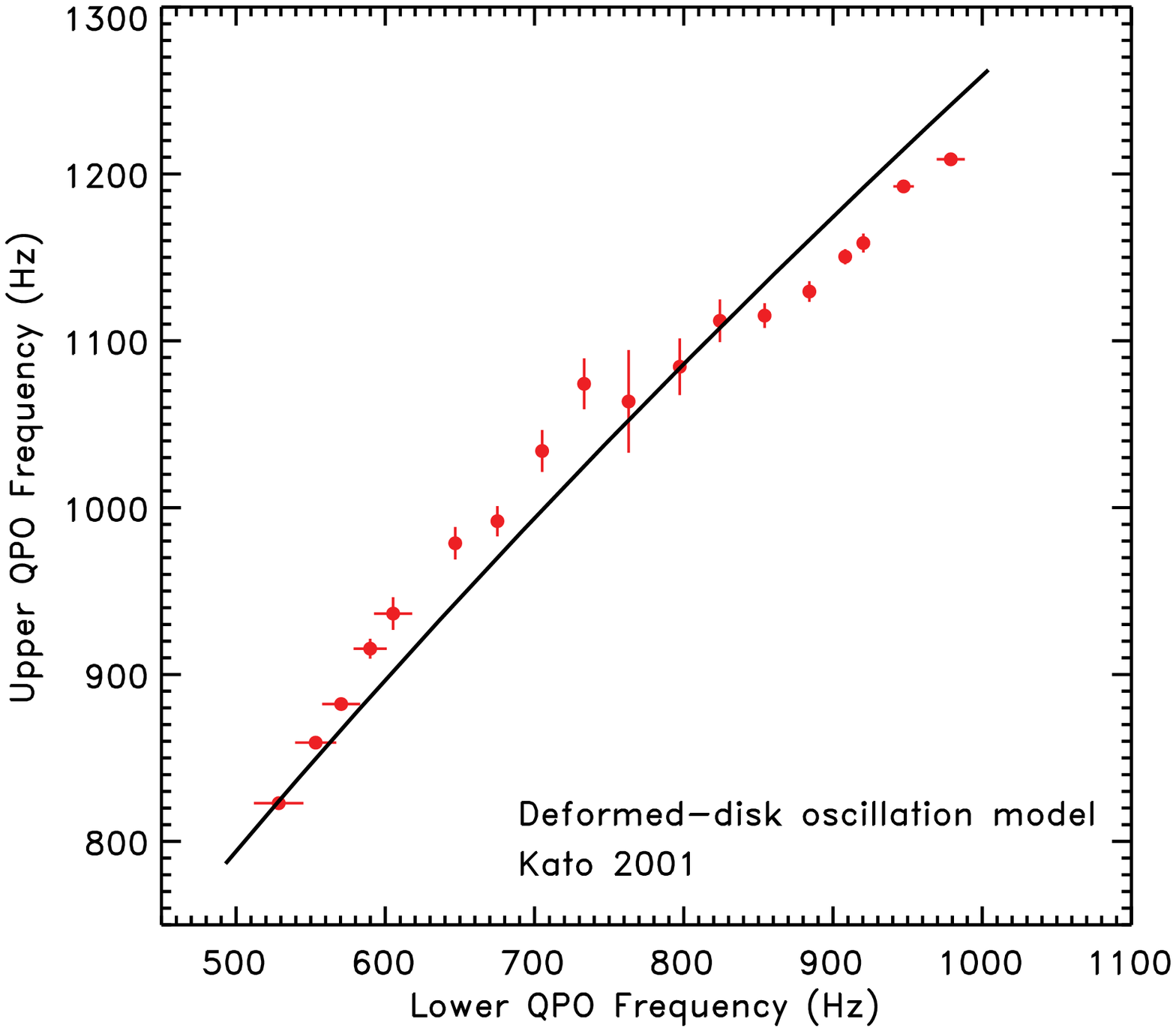}
\includegraphics[width=80mm]{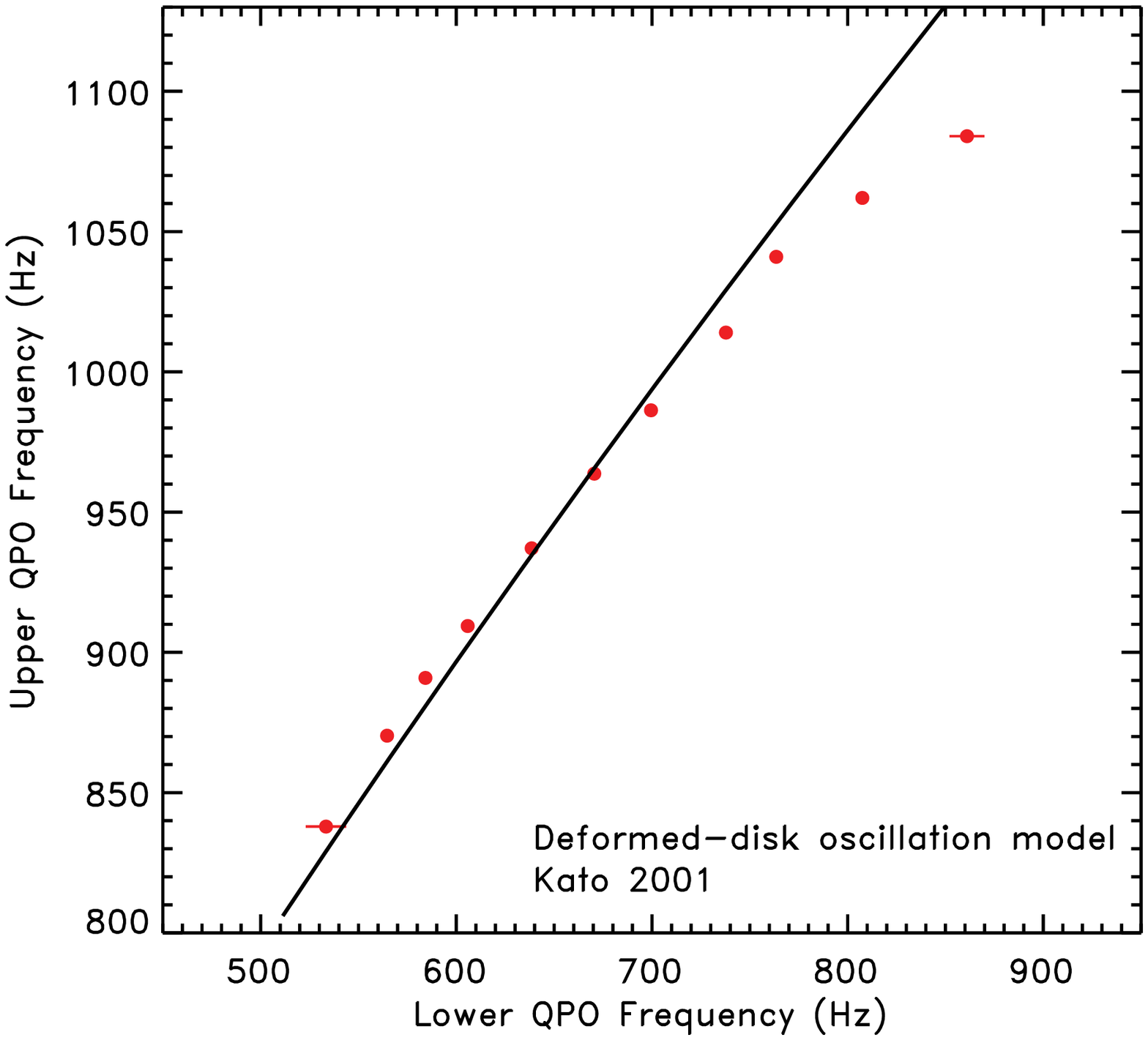}
\caption{The fitting results to the deformed-disk resonance model for 4U 1636-53 (top) and Sco X-1(bottom). The
solid curve represents the best fitting result with $M\approx2.4$ $M_\odot$, $j=0$. The predicted $M$ and $j$
are almost identical in the two LMXBs. }\label{fig:deformed}
\end{figure}

Figure \ref{fig:deformed} exhibits the best fittings for 4U 1636-53 and Sco X-1. The fitting results
($M\approx2.4$ $M_\odot$, $j=0$) are consistent with that in \citet{kato07}. The model gives a large NS mass. At
the same time, the fittings to the two different NS systems give the same set of $M$ and $j$. However the model
describes the measured data points relatively better than most of other models, though it predicts $\nu_2$
slightly larger than that observed at high frequencies.

\subsubsection{The `-1r, -2v' resonance model}

Unlike the deformed disk oscillation model, the perturbations in the `-1r, -2v' resonance model \citep{torok07,
bakala08} are not stressed in the azimuthal direction. In this model, the kHz QPOs are excited by the resonance
between the radial $m=1$ and the vertical $m=2$ modes. The excited QPOs are supposed to be,
\begin{equation}
\nu_1=\nu_{\rm K}-\nu_r,
\end{equation}
\begin{equation}
\nu_2=2\nu_{\rm K}-\nu_\theta.
\end{equation}

\begin{figure}[!htp]
\includegraphics[width=80mm]{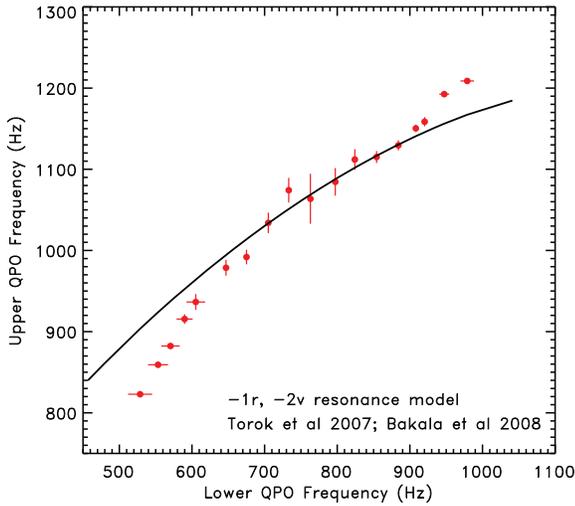}
\includegraphics[width=80mm]{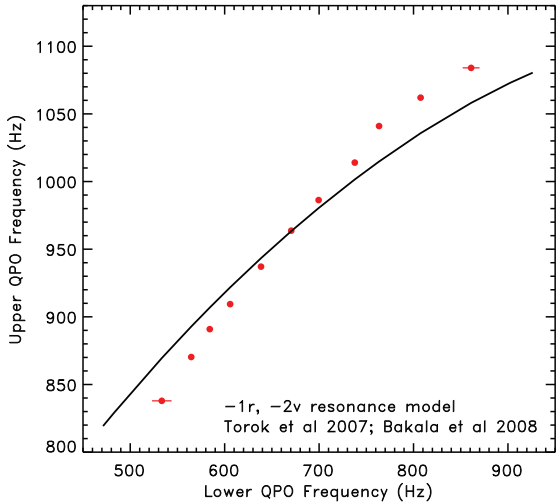}
\caption{The fitting results to the `-1r, -2v' resonance model for 4U 1636-53 (top) and Sco
X-1(bottom).}\label{fig:-1r-2v}
\end{figure}

The fitting results to the model are displayed in Figure \ref{fig:-1r-2v}. Like the precession and forced $1:3$
resonance models, the model predictions cannot reproduce the observations, especially at low and high
frequencies. The fittings have very large $\chi^2$ and give $M=2.4$ $M_\odot$ reaching the upper limit in the
fitting.

\subsection{Comparison with the higher-order nonlinearity model}

\citet{muk09} treated accreting systems as damped harmonic oscillators. These oscillators exhibit epicyclic
oscillations with higher-order nonlinear resonance. The resonance is expected to be driven by the coupling
between the strong disturbance from a NS and the weaker one from the flow. In the model, the lower and upper kHz
QPOs are proposed to be,
\begin{equation}
\nu_1=\nu_\theta-\frac{\nu_{\rm s}}{2} \label{hnm_nu1},
\end{equation}
\begin{equation}
\nu_2=\nu_r+\frac{n}{2}\nu_{\rm s} \label{hnm_nu2}.
\end{equation}

In the disk around a NS, $n=1$ corresponds to a nonlinear coupling, resulting in
$\Delta\nu=\nu_2-\nu_1\sim\nu_{\rm s}/2$; whereas $n=2$ corresponds to a linear coupling, resulting in
$\Delta\nu\sim\nu_{\rm s}$. However, $n=3$ may correspond to the higher-order coupling which is expected to be
too weak to produce any observable effects. For $n=1$ and $n=2$, the model divides NSs into fast and slow
rotators.

To compute the QPO frequencies, the spin parameter $j$ should be determined in the following way. If a NS is
considered to be spherical in shape with equatorial radius $R$, spin $\nu_{\rm s}$, mass $M$, radius of
gyration $R_G$, then the moment of inertia and the spin parameter are computed by,
\begin{equation}
I=MR_G^2,    \qquad j=\frac{I\Omega_s}{GM^2/c}, \label{hmn_j}
\end{equation}
where $\Omega_s=2\pi\nu_{\rm s}$. It is known that for a solid sphere $R_G^2=2R^2/5$ and for a hollow sphere
$R_G^2=2R^2/3$. However a very fast rotating NS is expected to be ellipsoidal and not completely solid.
Therefore $0.35\le(R_G/R)^2\le0.5$ is chosen in the model.

\begin{figure}[!htp]
\includegraphics[width=80mm]{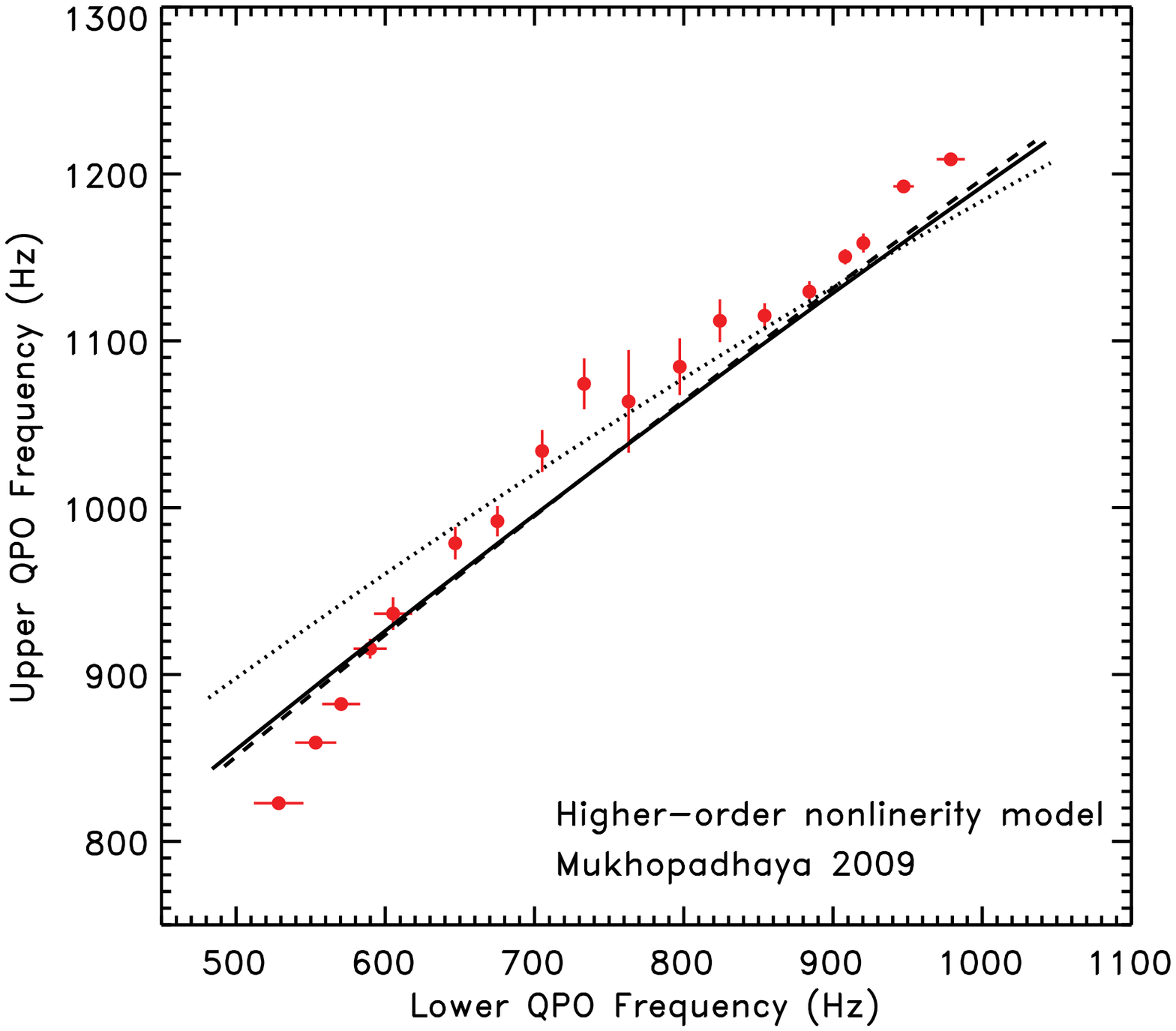}
\includegraphics[width=80mm]{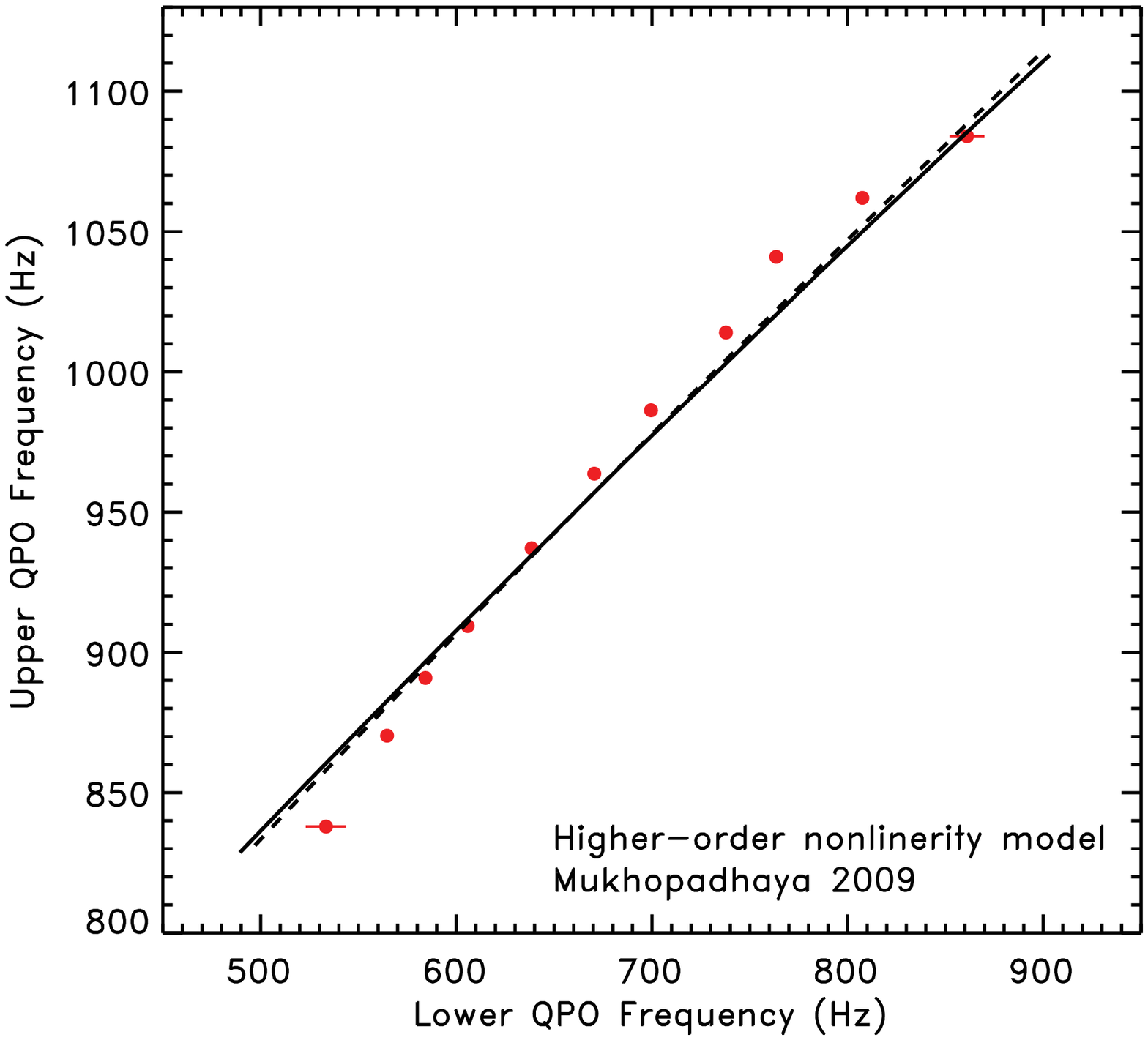}
\caption{The fitting results to the higher-order nonlinearity model for 4U 1636-53 (top) and Sco X-1(bottom).
Top panel: the solid curve represents the best fitting result under $n=1$. The dashed curve, which almost
overlaps the solid one, is the fitting under $n=2$. The dotted curve and dot-dashed one denote the fittings by
setting $\nu_{\rm s}=581$ Hz under $n=1$ and $n=2$, respectively. Bottom panel: the solid and dashed curves
indicate the best fitting results under $n=1$ and $n=2$.}\label{fig:hnm}
\end{figure}

The model parameters, i.e. $M$, $\nu_{\rm s}$, $R$ and $(R_G/R)^2$, can be obtained by the fitting. For
4U 1636-53, \citet{muk09} treated it as a fast rotator with $\nu_{\rm s}=581$ Hz and fit the frequency relation
under $n=1$. In his work, only six data points were collected in the diagram of $\Delta\nu$ versus $\nu_1$. Then
he excluded the data points at low frequencies, corresponding to the ones with large deviations from the model.
Finally he gave a low NS mass about $1.2\sim1.4$ $M_\odot$. For Sco X-1, the fitting was done both under $n=1$
and $n=2$ and he argued that Sco X-1 is a slow rotator with $n=2$ and $\nu_{\rm s}$ about $300$ Hz.

Our fitting results with different $n$ are shown in Figure \ref{fig:hnm}. Here we fit all the data points in the
two NS systems without any exclusion. For 4U 1636-53, the best fitting value is $M=1.40$ $M_\odot$,
$\nu_{\rm s}=489$ Hz, $R=23.1$ km, $(R_G/R)^2=0.38$ under $n=1$ and $M=1.40$, $\nu_{\rm s}=307$ Hz, $R=26.5$ km,
$(R_G/R)^2=0.46$ under $n=2$. We find that under $n=1$ and $n=2$ the model gives the curves almost superposed
(solid and dashed in the figure). Moreover, $\nu_2$ predicted by the model is too high at low frequencies. By
setting $\nu_{\rm s}=581$ Hz and $n=1$, we obtain the results with much larger deviations (dotted curve). For
Sco X-1, the model curves under $n=1$ and $n=2$ almost overlap, resulting in $467$ Hz and $294$ Hz spin
frequencies. The best fitting value of other parameters is $M=1.40$ $M_\odot$, $R=23.0$ km, $(R_G/R)^2=0.40$
under $n=1$ and $M=1.40$, $R=27.3$ km, $(R_G/R)^2=0.45$ under $n=2$. For $n=2$, our results are consistent with
that in \citet{muk09}.

\subsection{Comparison with the tidal disruption model}

Tidal disruption of the orbits of low-mass satellites around a Schwarzschild black hole has recently been
studied by \citet{cadez08}. In the clumps of material orbiting such a black hole, a spherical blob can be
squeezed and stretched by tidal forces into a ring-like shape along the orbit, and thus making radial
oscillations \citep{cadez09}. With simulations of such accretion processes, they generated simulated light
curves and fit the power spectra of the light curves. Both twin kHz QPOs are found and the peak frequencies are
supposed to be,
\begin{equation}
\nu_1=\nu_{\rm K},
\end{equation}
\begin{equation}
\nu_2=\nu_{\rm K}+\nu_r.
\end{equation}

\begin{figure}[!htp]
\includegraphics[width=80mm]{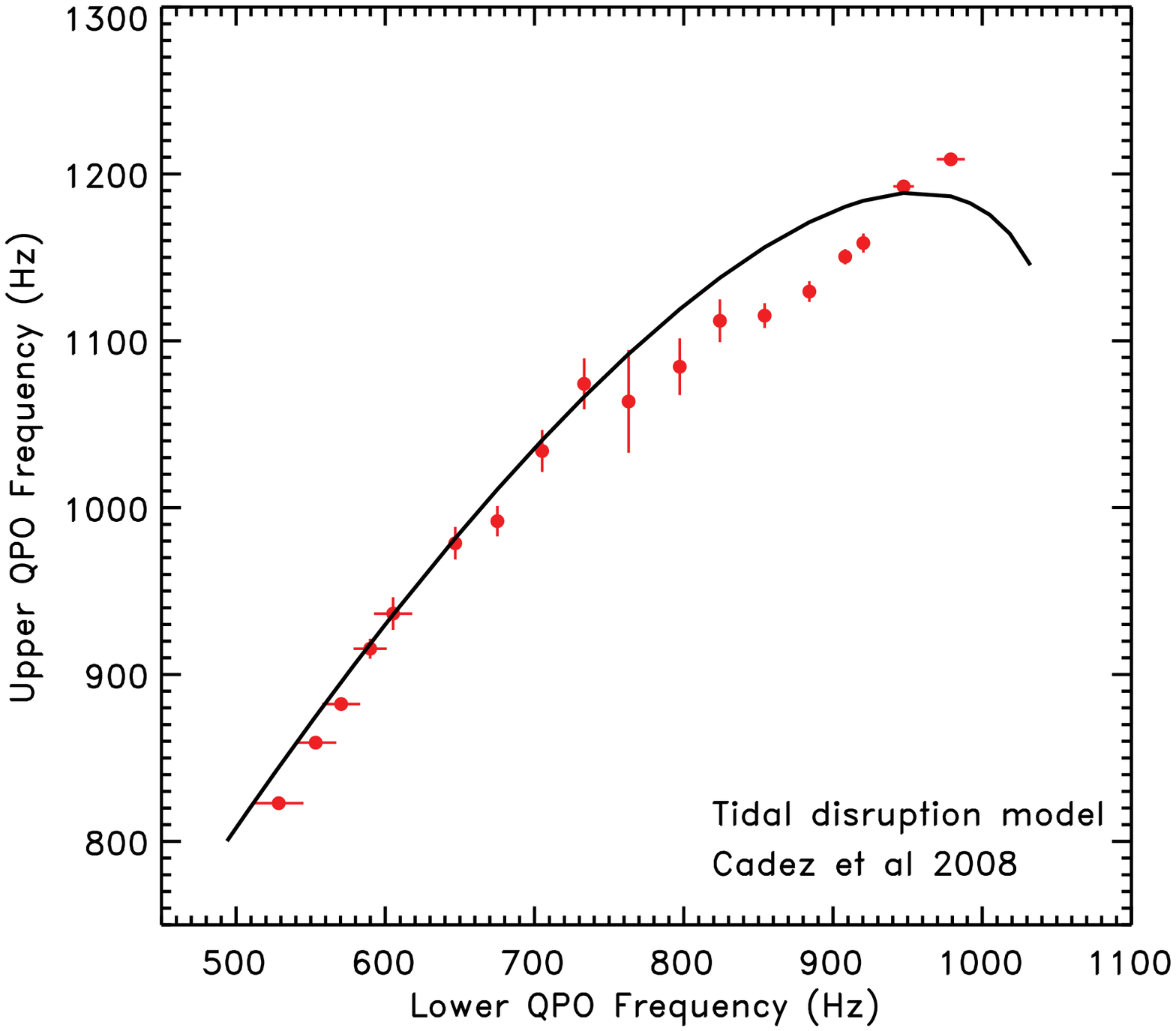}
\includegraphics[width=80mm]{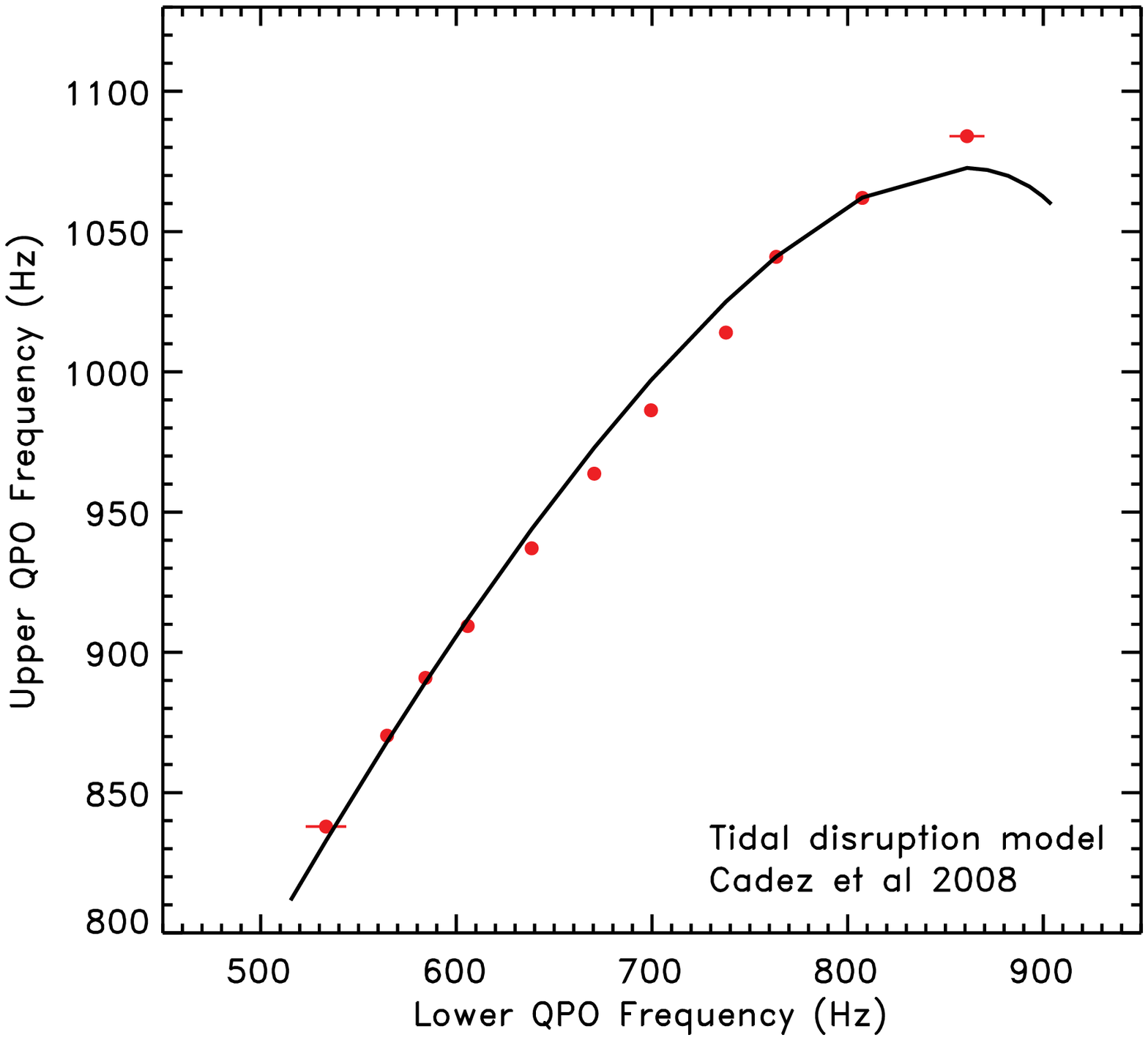}
\caption{The fitting results to the tidal disruption model for 4U 1636-53 (top) and Sco X-1 (bottom).
The curves exhibits the disagreement with the data points when $\nu_1>800$ Hz.} \label{fig:tidal}
\end{figure}

Figure \ref{fig:tidal} shows the best fittings to this model in the two NS systems. As we see, the model
describes well the main parts of the frequency relations, particularly at low frequencies ($\nu_1\le800$ Hz). At
high frequencies, however, the model predicts maximum value of $\nu_2$ and then a sharp decrease. It is not
supported by observations. Another incompatibility is a high NS mass predicted, which is up to $2.4$ $M_\odot$
in this model.

\subsection{The Rayleigh-Taylor gravity wave model}

\citet{ot99} and \citet{titar03} described QPOs by the Rayleigh-Taylor instability associated with Rossby waves
and rotational splitting. Twin kHz QPOs are explained as oscillations of large scale inhomogeneities (hot blobs)
thrown into the NS's magnetosphere. Participating in the radial oscillations with the Keplerian frequency
$\nu_{\rm K}$, such blobs are also simultaneously under the influence of the Coriolis force. For such mode of
oscillations, $\nu_2$ and $\nu_{\rm K}$ hold an upper hybrid frequency relation: $\nu_2^2-\nu_{\rm
K}^2=4\nu_{\rm m}^2$, where $\nu_{\rm m}$ is the rotational frequency of the magnetosphere near the
equatorial plane. If the magnetosphere corotates with the NS (solid-body rotation), then the spin rotation of
the NS would be determined. For the first order approximation, $\nu_{\rm m}=\nu_{\rm s}=const$. Within the
second-order approximation, the slow variation of $\nu_{\rm m}$ as a function of $\nu_{\rm K}$ reveals the
structure of the magnetospheric differential rotation. Hence, in the model,
\begin{equation}
\nu_1=\nu_{\rm K}
\end{equation}
\begin{equation}
\nu_2=(\nu_{\rm K}^2+4\nu_{\rm m}^2)^{1/2},
\end{equation}
and within the dipole-quadrupole-octupole approximation of the magnetic field, the rotation frequency of the
magnetosphere is,
\begin{equation}
\nu_{\rm m}(\nu_{\rm K})=C_0+C_1\nu_{\rm K}^{4/3}+C_2\nu_{\rm K}^{8/3}+C_3\nu_{\rm K}^4,
\end{equation}
where $C_0=\nu_{\rm s}$, $C_2=2\sqrt{C_1C_3}$.

\begin{figure}[!htp]
\includegraphics[width=80mm]{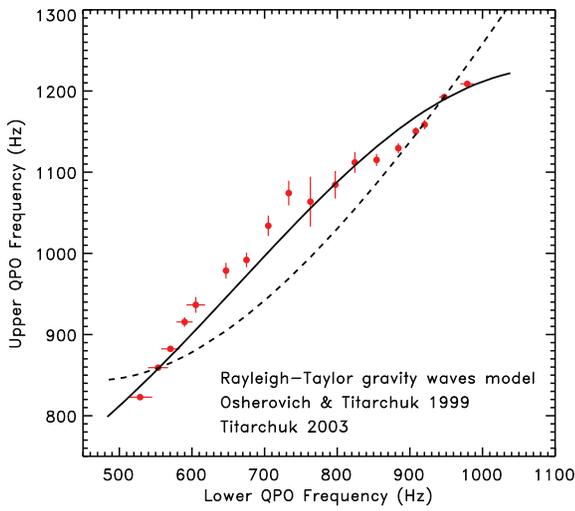}
\includegraphics[width=80mm]{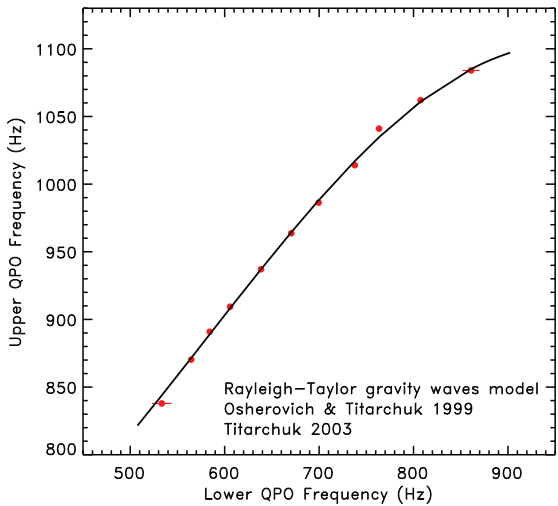}
\caption{The fitting results to the Rayleigh-Taylor gravity wave model for 4U 1636-53 (top) and Sco X-1
(bottom). The solid curves are the best fitting results while the dash curve in the top panel is the fitting
result by setting $\nu_{\rm s}=581$ Hz. }\label{fig:RTGwave}
\end{figure}

Our fitting treats $M$, $C_0$, $C_1$, $C_3$ as free parameters. The fitting results are shown in Figure
\ref{fig:RTGwave}. For 4U 1636-53, the best fitting (solid curve) returns $M=1.78$ $M_\odot$, $C_0=371$,
$C_1=-0.050$, $C_3=-7.7$. The spin frequency $\nu_{\rm s}=371$ Hz, not close to $581$ Hz. When we fix $\nu_{\rm
s}=581$ Hz, a concave curve (dashed) is obtained, whereas the track of data points bends downward with a convex
shape. For Sco X-1, the best fitting values are $M=1.66$ $M_\odot$, $C_0=350$, $C_1=-0.046$, $C_3=-10.5$,
respectively. Notice that the NS spin frequency of $350$ Hz is consistent with $345$ Hz in \citet{ot99}.
Moreover, the resulting NS mass $1.66$ $M_\odot$ is reasonable based on the EOS of NS \citep{lattimer07}.

\subsection{Comparisons with the models including the effect of magnetohydrodynamics}

The last two models above have included the effect of magnetohydrodynamics (MHD) around a rotating NS.
In a LMXB containing a magnetized NS, the material in the accretion disk first rotates in a Keplerian motion,
then corotates with the magnetosphere as it is trapped by the NS magnetic field at the magnetospheric radius,
and finally flows along the field lines to the polar cap of the NS. Some resonant modes may be excited by the
perturbations at the magnetospheric radius \citep{zhang04, shi09}.

\subsubsection{The MHD Alfv\'en wave oscillation model}\label{se:zhang}
%%(AWOM; Zhang 2004)

\citet{zhang04} explained the twin kHz QPOs with the MHD Alfv\'en wave oscillations excited by the distortion of
the NS magnetosphere. The model assumes that the infalling MHD material of the Keplerian accretion flow distorts
the magnetosphere in the regions with enhanced mass density gradients, leading to resonant shear Alfv\'en waves.
In this model, the upper frequency QPO is the Keplerian orbital frequency,
\begin{equation}
\nu_2=\nu_{\rm K}=1850AX^{3/2} \quad {\rm Hz},
 \label{eq:AMO_nu2}
\end{equation}
with the parameters $X=R/r$ and $A=(m/R_6^3)^{1/2}$, where $R_6=R/10^6$ $({\rm cm})$ and $m=M/M_\odot$ are the
NS radius $R$ and mass $M$ in units of $10^6$ cm and solar masses, respectively. The quantity $A^2$ is
proportional to the average mass density of the NS, expressed as,
$\langle{\rho}\rangle=3M/(4\pi{R^3})\approx{2.4}\times10^{14}({\rm g}/{\rm cm}^3)(A/0.7)^2$.

The lower frequency QPO is identified as the Alfv\'en oscillation frequency, given as,
\begin{equation}
\nu_1=\nu_2X^{3/4}\sqrt{1-\sqrt{1-X}}. \label{eq:AMO_nu1}
\end{equation}

\begin{figure}[!htp]
\includegraphics[width=80mm]{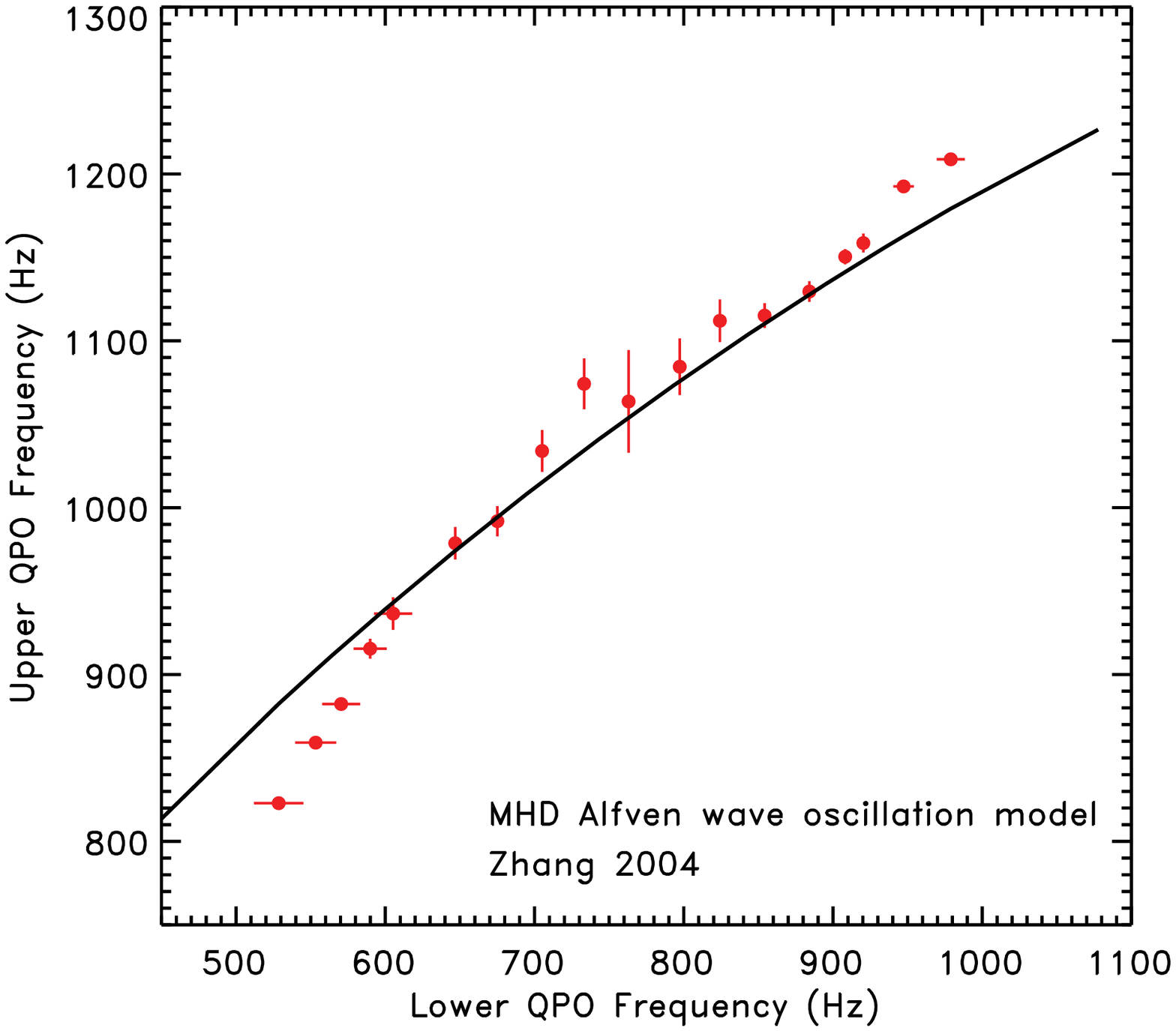}
\includegraphics[width=80mm]{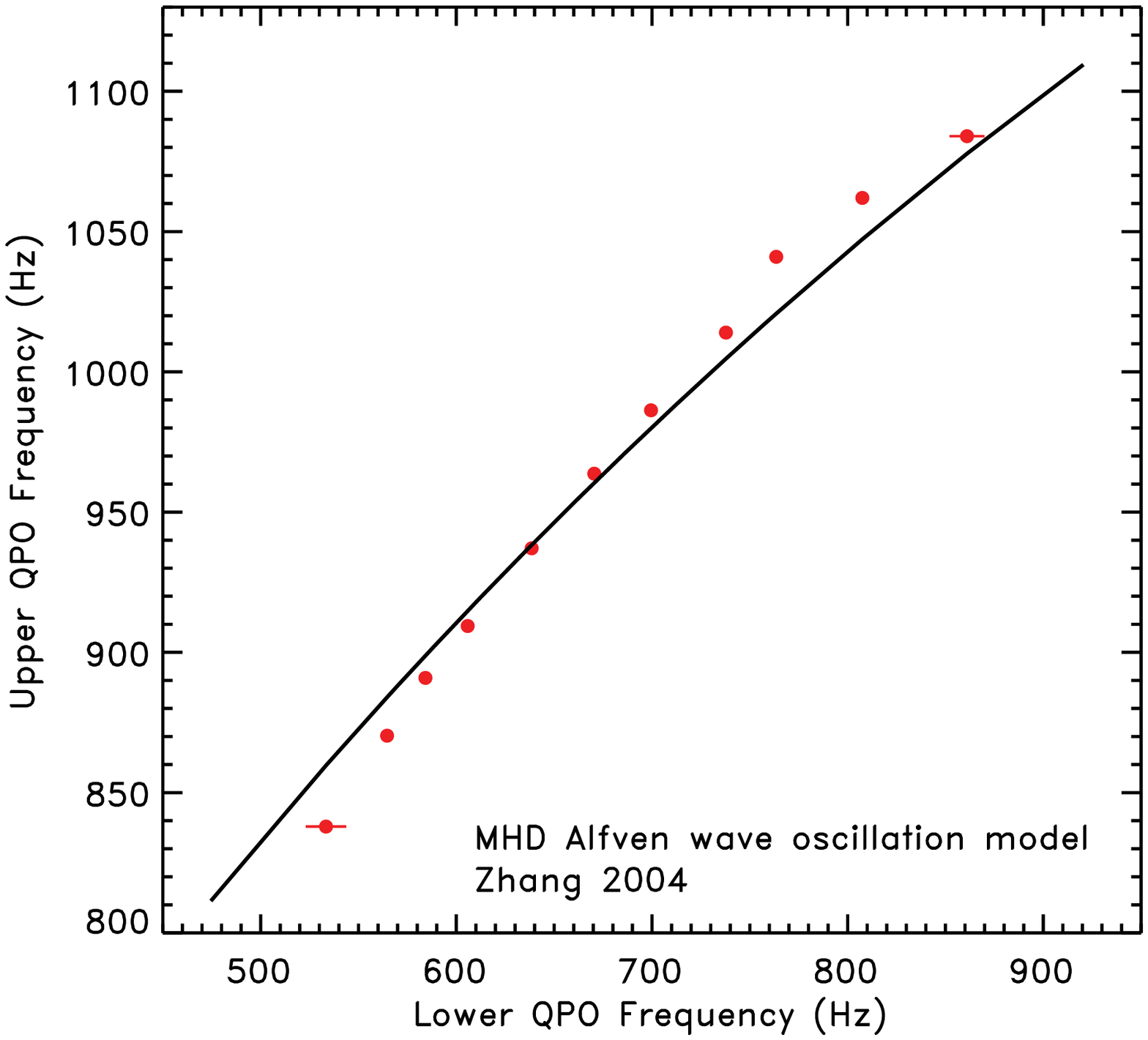}
\caption{The fitting results to the MHD Alfv\'en wave model for 4U 1636-53 (top) and Sco X-1(bottom).
}\label{fig:Alfven}
\end{figure}

Since $X$ is eliminated in the fitting process, we only have one parameter $A$. The comparisons between the
model predictions and the observations are shown in Figure \ref{fig:Alfven}. Similar to the precession models,
this model also predicts $\nu_2$ too high at low frequencies and too low at high frequencies, leading to the
increased deviations from the observations at low and high frequencies. The result also indicates that
$\Delta\nu$ predicted by this model decreases too sharply compared to the observations. The fitting for Sco X-1
is somewhat better than that for 4U 1636-53. Our result of $A\approx0.7$ agrees with that obtained by
\citet{zhang08}, in which the relation of $\Delta\nu$ versus $\nu_2$ was fitted and the result shows a
discrepancy with the observations.

\subsubsection{The MHD model}
%Shi, C, Li, X, The magnetohydrodynamics model of twin kilohertz quasi-periodic oscillations in low-mass X-ray binaries, MNRAS, 392, 264, 2009

\citet{shi09} presented another explanation for kHz QPO signals in LMXBs based on MHD oscillation modes in a
NS's magnetosphere. Several MHD wave modes are derived by solving the dispersion equations. They proposed that
the coupling of the two resonant MHD modes may lead to the twin kHz QPOs. Finally, they presented the following
linear frequency relations,

\begin{equation}\label{eq:lscs}
\nu_2=\sqrt{1+\delta^2}(\nu_1+\nu_{\rm s}),                             \qquad ({\rm LSCS})
\end{equation}
\begin{equation}
\nu_2=\frac{1}{\sqrt{1+\varepsilon^2}}(\nu_1+\nu_{\rm s}),    \qquad ({\rm SSCS})
\end{equation}
where $\delta^2=(\lambda^2-\eta^2)/(1+\eta^2)$ and $\varepsilon^2=(\eta^2-\lambda^2)/(1+\lambda^2)$.
Here $\lambda$ and $\eta$ are two constants linking the Alfv\'en velocity, acoustic velocity and Keplerian
velocity of MHD wave in the model. The model divided the twin kHz QPOs into two groups with the slope of
$\nu_2/\nu_{\rm s}$ vs. $\nu_1/\nu_{\rm s}$ relation either larger or smaller than 1.0, i.e., the large slope
coefficient sources (LSCS) and the small slope coefficient sources (SSCS), respectively. With our fitting, we
find that both 4U 1636-53 and Sco X-1 belong to the latter group, since Eq.~(\ref{eq:lscs}) gives bad fitting
results with reduced $\chi^2>10^5$.

\begin{figure}[!htp]
\includegraphics[width=80mm]{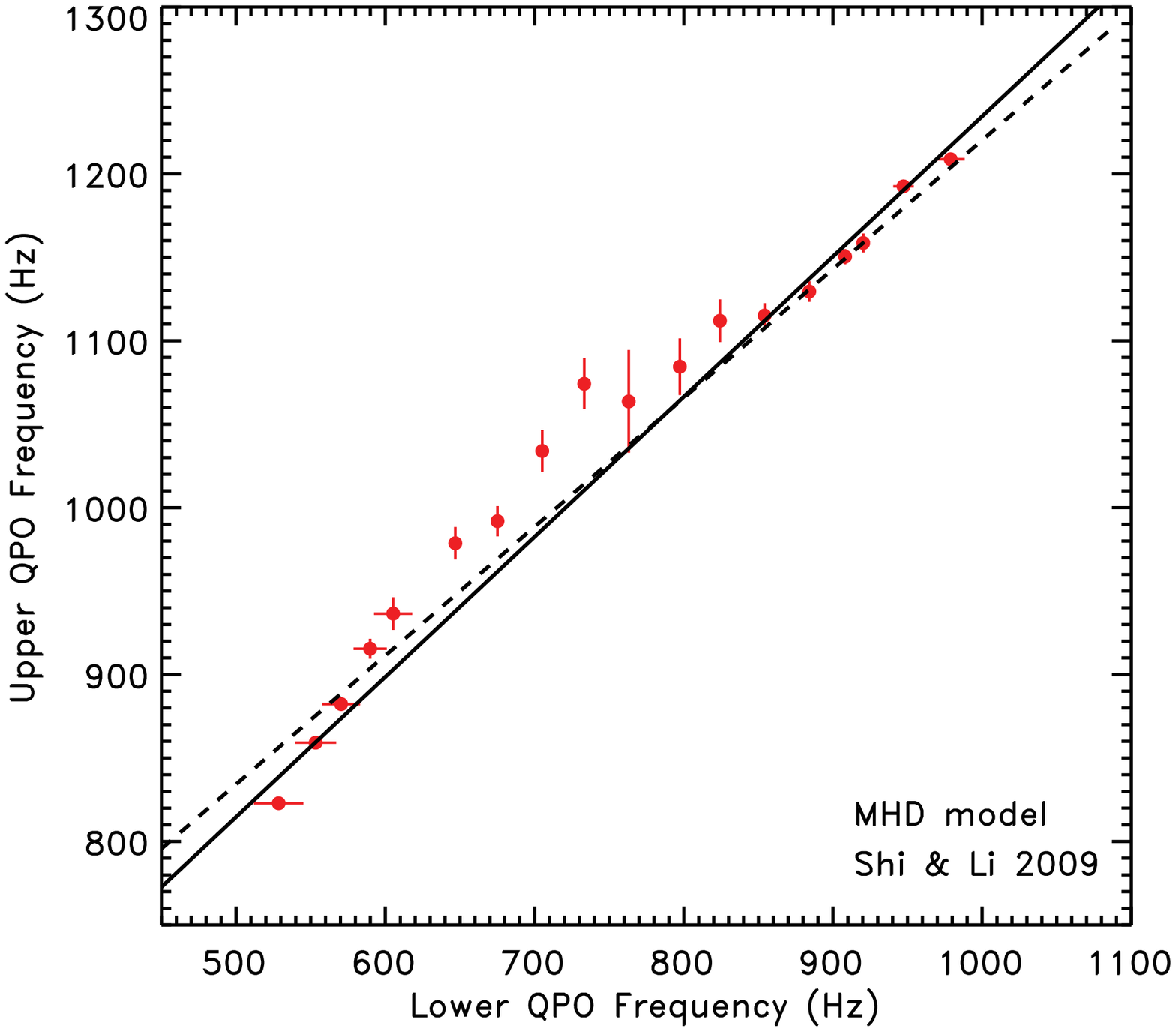}
\includegraphics[width=80mm]{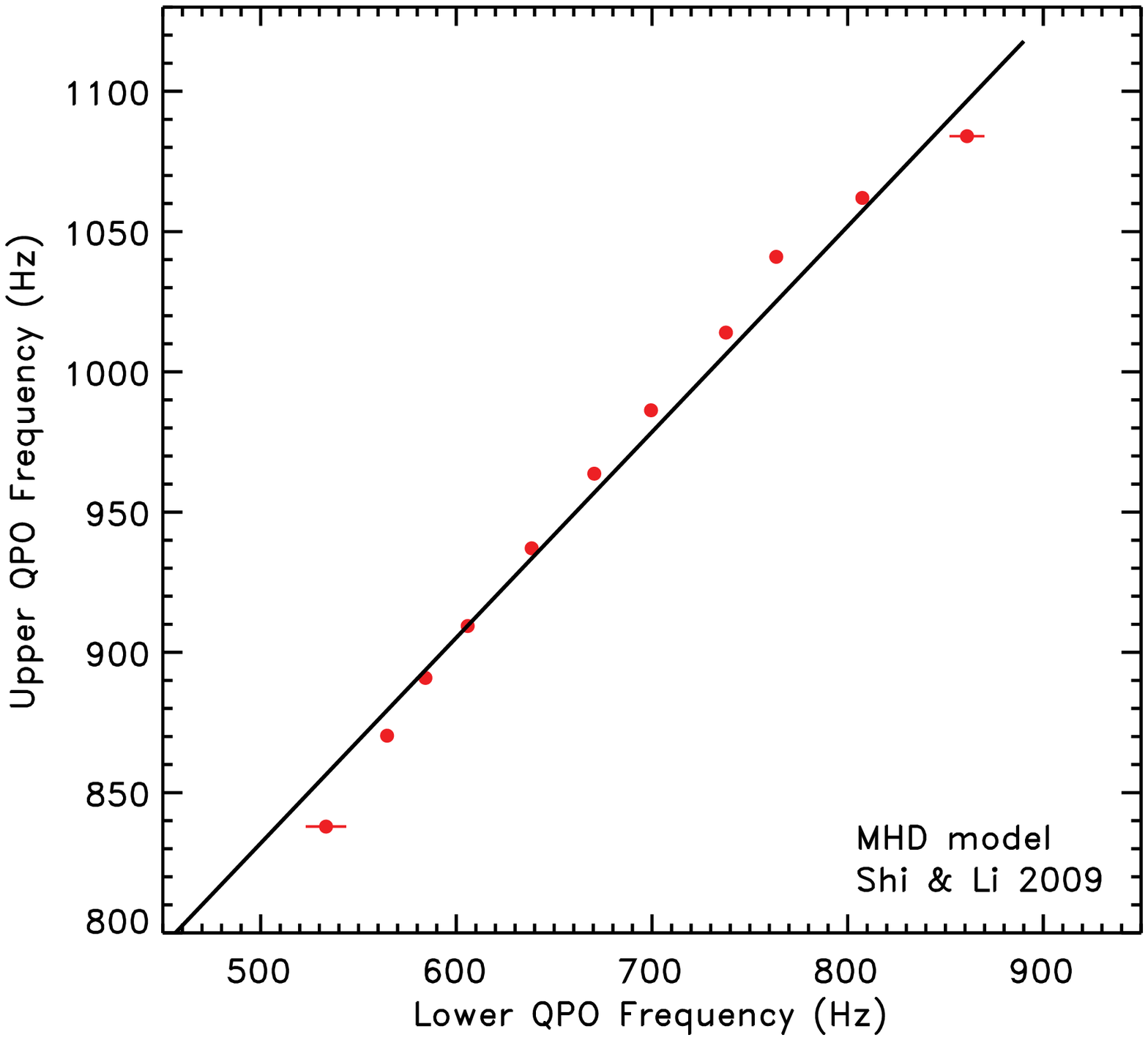}
\caption{The fitting results to the MHD model for 4U 1636-53 (top) and Sco X-1(bottom). The solid curves are the
best fitting results. In the top panel, the dashed curve is the result with $\nu_{\rm s}=581$ Hz
fixed.}\label{fig:mhd}
\end{figure}

The fitting results are plotted in Figure \ref{fig:mhd}. Firstly, both for 4U 1636-53 and Sco X-1, the observed
relations of $\nu_2$ with $\nu_1$ are not linear; the data points have a track to bend downwards. Secondly, for
4U 1636-53, the best fitting predicts $\epsilon=0.65$ and $\nu_{\rm s}=470$ Hz. The spin frequency is
not close to $581$ Hz. Also, \citet{shi09} cannot fit well 4U 1636-53 by holding $\nu_{\rm s}=581$ Hz. Their
result gives $\epsilon=0.77$ and a large reduced $\chi^2=23$. In the case of Sco X-1, our fitting gives
$\nu_{\rm s}=525$ Hz, considerably larger than that from other models.

\section{Discussion and conclusion}\label{se:con}

We have presented the newly obtained and more accurate results on the frequency relations of the kHz twin QPOs
for 4U 1636-53 and Sco X-1. The peak frequencies of lower and upper QPOs are almost directly proportional to
each other. The data points tend to bend downwards between $500$ to $1250$ Hz in the diagram of $\nu_1$ versus
$\nu_2$. Both of the frequency relations show some subtle structure around $\nu_1\approx800$ Hz.

Based on the frequency relations, we have systematically investigated the predictive ability of all currently
available models, with which the frequency relation can be calculated. Our conclusions are as follows:

(1) The sonic-point and spin-resonance model seems to be only suitable for Z-sources. The model describes well
the frequency relation for Sco X-1. However, in the case of 4U 1636-53, the model predicts that $\Delta\nu$ is
always less than half of the spin frequency for a prograde flow, while the observed $\Delta\nu$ can be
larger and smaller than $\nu_{\rm s}/2$. It indicates that a single rotational direction of accretion
flow cannot explain the behavior of $\Delta\nu$. According to the model, such behavior of $\Delta\nu$ may be
produced by a flow that retrogrades when it is far away from the NS, but then switches to a prograde orbit at
some special radius when it is closer to the NS. However no evidence is found to support such sudden switch of
the orbital motion.

(2) The $2:3$ parametric resonance model predicts a frequency relation bending upwards. Within the limit of NS
parameters for reasonable NS EOS, it cannot give $\nu_1$ as high as observations. Further more, as claimed in
previous papers \citep{belloni05, zhang06}, this model leads to the predicted $\Delta\nu$ increase with the QPO
frequency. However, the observed downward-bending track in the diagram of $\nu_1$ versus $\nu_2$ indicates a
rough inverse proportion between $\Delta\nu$ and $\nu_1$ (or $\nu_2$). As regards the forced resonance model,
the $1:2$ model predicts a decrease of $\nu_2$ at high frequencies; the $1:3$ model predicts $\nu_2$ higher
(lower) than the observed values at low (high) frequencies. In fact, all of the orbital resonance models are
introduced to explain the observed clustering of $2:3$ ratios between $\nu_1$ and $\nu_2$. However,
\citet{belloni05} have demonstrated that a simple random walk of the QPO frequencies can reproduce qualitatively
the observed distributions in frequency and frequency ratio. Later \citet{bou09} have pointed out that the
clustering originates naturally from the sensitivity-limited observations and does not support preferred
frequency ratios in NS systems. Our results therefore suggest that the orbital resonance models should be
further investigated in order to improve their predictive power for the frequency relation.

(3) All the precession models nearly overlap with each other. Their predicted $\nu_2$ is higher (lower) than
that observed at low (high) frequencies. The deviations from the observations increases significantly at high
and low frequencies. In more detail, the relativistic and vertical precession models predict a NS mass higher
than $\sim2.2$ $M_\odot$. As can be found from Table 3, when we relax the fitting limits of $M\in[1.4, 2.4]$ and
$j\in[0, 0.3]$, the extended analysis shows that these two models could describe the data points better with
relatively higher $M$ and $j$. Essentially, the inferred high NS mass may be arisen from the assumption of the
vacuum circumstance around the NS in introducing the periastron precession term \citep{zhang09}. It should be
mentioned that considering a small eccentricity ($\lesssim 0.1$) which decreases with increasing $\nu_\phi$, the
relativistic precession model would explain the frequency relation better (Stella and Vietri 1999). In this
paper, we do not consider the effect of eccentricity on the frequency relation because we focus on the NS
properties. For the total precession model, it gives lower values of $M$ and $j$ but larger reduced $\chi^2$.

(4) The deformed-disk resonance model describes the observations relatively better than most of models. The
fittings give the same $M$ and $j$ for the two LMXBs. It suggests that the model may reveal some common
properties of Atoll and Z sources. Nevertheless the high mass predicted and the deviations at high frequencies
show the model could be modified. For example, the effect of magnetic field could be taken into account. For the
`-1r, -2v' resonance model, it behaves like the procession models. Considering that the best
fitting results of these two models approach the upper limit of $M$, we also performed the extended fitting.
Both for 4U 1636-53 and Sco X-1, the fitting results do not improve significantly.

(5) The higher-order nonlinearity model classifies NSs based on the values of $n$. After the investigation, one
can notice that the model predict the nearly identical frequency relations under all values of $n$ taken here.
Thereby, given that we do not know the spin frequency of Sco X-1, the classification that Sco X-1 is a slow
rotator with $n=2$ is not well founded. The superposed fitting curves under $n=1$ and $n=2$ for 4U 1636-53 also
indicate some kinds of ambiguity of the classification. Apart from that, when $n=1$ is chosen, like that
in \citet{muk09}, the spin frequency predicted is not close to $581$ Hz. Besides, the model predicts a very low
NS mass reaching the lower limit in our fitting. Our extended analysis shows that the NS masses
for the two sources are predicted down to $1.0$ $M_\odot$, quite low for most known NS EOS.

(6) The tidal disruption model can describe the main part of the observed frequency relations, though it
predicts a high NS mass. Maybe it is due to the essential difference between NS and black hole systems. This
implies that the kHz QPOs should be greatly affected by the surroundings close to the central object. After all,
the agreement with the observations at low frequencies is remarkable, corresponding to the place relatively
distant to the center. At that place, the clumps of material (or particles) are not being exposed to the compact
object so much. As can be found from Table 3, the fitting results of the model can be improved
greatly in our extended fitting, in favor of very large NS masses. The model should be modified to better
describe the data points at high frequencies and to get a more reasonable mass for NS.

(7) The Rayleigh-Taylor gravity wave model can follow the frequency relation in Sco X-1, while for 4U
1636-53 it cannot predict the $581$ Hz spin frequency. This may be because the dipole-quadrupole-octupole
approximation is not sufficiently accurate for the magnetic field. Therefore, this model seems promising for
explaining the origin of kHz QPOs if its description of NS magnetic field is more accurate. For the sake
of comprehensiveness, it should be noted that this model predicts not only the high-frequency QPOs, but also the
low frequency ones. When the low QPOs are also considered, the fittings do not always work, unless for one
particular frequency range one of the low-frequency QPOs is assumed to be a harmonic of an unseen one, whereas
in the other intervals it is the fundamental frequency.

(8) The MHD Alfv\'en wave oscillation model has the same problem as the precession models with increased
deviations from the observations at high and low frequencies. It should be noted that the model was put forward
based on the analogy of the solar coronal atmosphere to a NS system. Though the solar coronal atmosphere has
been studied a lot, the mechanism of Alfv\'en wave oscillations in a NS system remains unclear. The model's
performance in our fittings indicates that such mechanism should be investigated further.

(9) The MHD model predicts a linear frequency relation, which is inconsistent with the measured frequency
relations. At the same time, the model cannot predict reasonable spin frequencies for the two NSs.

Generally, we also find that:

(10) These models diverge strongly in their predictions of the NS properties. Different models predict spin
frequency from less than $300$ Hz to more than $600$ Hz. The angular momentum is predicted as from 0 to
0.3, covering entirely the range of the limit in the fitting. The predicted NS mass from different models also
covers the whole range $[1.4, 2.4]$ $M_\odot$.

(11) The problem of increased deviations at high and low frequencies exists in six models: the forced $1:3$
model, the three precession models, the `-1r, -2v' resonance model and the MHD Alfv\'en wave oscillation model.
The first five models almost overlap in the plot of their fitting results. Since $\nu_{\rm
K}\approx\nu_\theta$ (exactly equal if $\nu_{\rm s}=0$), these five models have nearly identical expressions of
$\nu_1$ and $\nu_2$. Those five models form the group of the largest $\chi^2$ in Table 2. The MHD Alfv\'en
oscillation model performs slightly better than those five models, despite that its $\chi^2$ remains much larger
than the other remaining models. All of the six models propose that the upper frequency QPO is Keplerian, i.e.
$\nu_2=\nu_{\rm K}$. It infers that for these models, the interpretation is not favored by the data.

(12) Those models including the effects of magnetic field obtain the best fitting results, such as the
sonic-point beat frequency and the Rayleigh-Taylor gravity wave model. At least, they can depict the frequency
relation for Sco X-1.

Finally, one should notice the fact that no model gives a statistically acceptable $\chi^2$ in the fittings. We
argue that all the models predicting a linear, power-law or any other frequency relation are not
fully supported by the observations, at least for this two sources.

After the investigation, we comment that since among these models we investigated here, three models of them
(deformed-disk resonance, tidal disruption and the Rayleigh-Taylor gravity wave model) have performed relatively
better than other models, we speculate that a model which combines these three models together could reveal the
physical origin of the observed kHz QPO signals. It is worth noticing that each of them still has its own
problems.

\acknowledgments SNZ acknowledges partial funding supports by Directional Research Project of Chinese Academy of
Sciences under project No. KJCX2-YW-T03, the National Natural Science Foundation of China under grant Nos.
10821061, 10733010, 10725313, and 973 Program of China under grant 2009CB824800.

%\clearpage
\begin{table}[!h]

\begin{minipage}{175mm}
\begin{center}
\caption{The fitting parameters of all models for 4U 1636-536 and Sco X-1. The errors have been computed by
setting $\Delta\chi^2=1$.}
\begin{tabular}{@{}ccccccccc@{}}
\hline
    &  \multicolumn{4}{c}{4U 1636-536}  & \multicolumn{4}{c}{Sco X-1} \\

\hline
Models             &  M ($M_\odot$) & j or spin (Hz) & Reduced $\chi^2$ & Dof  &M ($M_\odot$) & j or spin (Hz) & Reduced $\chi^2$ & Dof\\
\hline
SP SR Case 1                 & - & -  & - &- &1.80$\pm$0.02 &  330$+2 \atop -1$  & 3.0 & 8\\
SP SR Case 2                 &1.545$+0.046 \atop -0.042$  & 652$\pm$6    & 7.5 &15 &1.80$\pm$0.02 &  659$+4 \atop -2$  & 3.0 & 8\\
fix spin             &1.746$+0.008 \atop -0.009$  & 581         &  $18.9$ &15 \\
\hline
Para. Res.      & great     &departure  &from &frequency &relation \\
linear relation &  2.04$\pm$0.01                    &0.19$\pm$0.01                                        & 11   & 14    & 2.09$\pm$0.01                 & 0.07$\pm$0.01                &71 & 7 \\
\hline
Forced 1:3   &  1.815$+0.003 \atop -0.004$                    &0.0001$+0.0010 \atop -0.0001$  & 186  & 16  & 1.97$\pm$0.01            & 0.000$+0.001 \atop -0.0$             &306 & 9 \\
Forced 1:2    &  2.095$+0.006\atop -0.003$                    &0.0$+0.005 \atop -0.0$              & 28    & 16      & 2.32$\pm$0.01          & 0.0$+0.001 \atop -0.0$        &54 & 9 \\
\hline
Rel. Pre.   &  2.319$+0.003 \atop -0.005$                    &0.30$+0.0 \atop -0.001$                 & 156  & 16      & 2.40$+0.0 \atop -0.01$                   & 0.250$\pm$0.001                &244 & 9 \\
Ver. Pre.   &  2.160$+0.003 \atop -0.004$                    &0.300$+0.0 \atop -0.001$               & 158   & 16     & 2.33$\pm$0.01              & 0.299$\pm$0.001                &230 & 9 \\
Tot. Pre.   &  1.814$+0.004 \atop -0.003$               &0.0$+0.001 \atop -0.0$    & 186   & 16     & 1.971$+0.001 \atop -0.002$      & 0.0$+0.001 \atop -0.0$               &306 & 9 \\
\hline
Deformed-disk   &  2.400$+0.0 \atop -0.001$                    &0.00$+0.0002 \atop -0.0$              & 31  & 16       & 2.400$+0.0 \atop -0.001$               & 0.0$+0.001 \atop -0.0$                &447 & 9 \\
\hline
`-1r, -2v' Res.             &  2.400$+0.0 \atop -0.001$        &0.238$+0.002 \atop -0.001$            & 154   & 16     & 2.400$+0.0 \atop -0.002$                   & 0.172$+0.002 \atop -0.001$                 &248 & 9 \\
\hline
Hi. non-line. n=1    &  1.401$\pm0.001$                               &488.6$+0.01 \atop -0.07$             & 85   & 14      & 1.40$+0.01 \atop -0.0$                   & 467 $\pm$1                &86 & 7 \\
fix spin                    &  2.10$+0.01 \atop -0.02$                                   &581                & 211 & 15\\
Hi. non-line. n=2    &  1.401$\pm0.001$                                   &306.7$+0.01 \atop -0.07$             & 70  & 14       & 1.40$+0.01 \atop -0.0$                   & 294$\pm$1               &70 & 7 \\
\hline
Tidal Disruption    &  2.400$+0.0 \atop -0.003$                        &0.166$+0.001 \atop -0.002$               & 18 & 16     & 2.400$+0.0 \atop -0.002$              & 0.045$+0.001 \atop -0.002$                &43 & 9 \\
\hline
R-T G. Wave          &  1.78$+0.47 \atop -0.38$                  &371.1$+1.2 \atop -2.7$                    & 10  & 14  & 1.66$+0.34 \atop -0.26$                  & 350.4$+0.3 \atop -0.8$                 &4.4 & 7 \\
fix spin                    &  1.86$+0.38 \atop -0.40$                                     & 581             &  $38$ & 15\\
\hline
Alfv\'en Wave Res.  & A=0.699$\pm$0.002                &                          &81   &17        &A=0.658$\pm$0.001              &                           &68  &10 \\
\hline
MHD                       &$\epsilon=$0.647$+0.013 \atop -0.014$ &470.3$+7.3 \atop -7.7$                   &10 & 16          &$\epsilon$=0.928$+0.001 \atop -0.007$ &635$+6 \atop -5$                           &56 & 9 \\
fix spin                    &$\epsilon$=0.928$+0.002 \atop -0.003$&581                  &21 & 17 \\
\hline
\end{tabular}
\end{center}
\end{minipage}
\end{table}

\begin{table}
\begin{minipage}{175mm}
\begin{center}
\caption{The fitting parameters of the extended analysis for 4U 1636-536 and Sco X-1.}
\begin{tabular}{@{}ccccccccc@{}}
\hline
    &  \multicolumn{4}{c}{4U 1636-536}  & \multicolumn{4}{c}{Sco X-1} \\

\hline
Models             &  M ($M_\odot$) & j or spin (Hz) & Reduced $\chi^2$ & Dof  &M ($M_\odot$) & j or spin (Hz) & Reduced $\chi^2$ & Dof\\
\hline
Rel. Pre.   &  2.8                    &0.49               & 131  & 16      & 3.07                   & 0.499                &165 & 9 \\
Ver. Pre.   &  2.46                   &0.5               & 141   & 16     & 2.65                    & 0.5              &188 & 9 \\
\hline
Deformed-disk   &  2.48             &0.00                & 22  & 16       & 2.45         & 0.0            &395 & 9 \\
\hline
`-1r, -2v' Res. &  3.61             &0.500                & 129   & 16     & 3.75          & 0.471           &183 & 9 \\
\hline
Hi. non-line. n=1    &  1.0          &460               & 50   & 14      & 1.0           & 438                &61 & 7 \\
Hi. non-line. n=2    &  1.0          &283               & 34  & 14       & 1.0           & 282                &59 & 7 \\
\hline
Tidal Disruption    &  3.33          &0.483              & 8 & 16       & 2.70          & 0.19           &26 & 9 \\
\hline
\end{tabular}
\end{center}
\end{minipage}
\end{table}

\end{document}